\pdfoutput=1
\documentclass[%
	paper=A4,					
	twoside=true,				
	openright,					
	parskip=full,				
	chapterprefix=true,			
	11pt,						
	headings=normal,			
	bibliography=totoc,			
	listof=totoc,				
	titlepage=on,				
	captions=tableabove,		
	draft=false,				
  parskip=half,
]{scrreprt}%

\newcommand{\thesisTitle}{Non-linear Associative-Commutative Many-to-One Pattern Matching with Sequence Variables}
\newcommand{\thesisName}{Manuel Krebber}
\newcommand{\thesisSubject}{Master Thesis}
\newcommand{\thesisDate}{April 24, 2017}

\newcommand{\thesisFirstReviewer}{Prof. Paolo Bientinesi, Ph.D.}
\newcommand{\thesisFirstReviewerUniversity}{\protect{RWTH Aachen University}}
\newcommand{\thesisFirstReviewerDepartment}{AICES, HPAC Group}

\newcommand{\thesisSecondReviewer}{Dr. Diego Fabregat-Traver}
\newcommand{\thesisSecondReviewerUniversity}{\protect{RWTH Aachen University}}
\newcommand{\thesisSecondReviewerDepartment}{AICES, HPAC Group}

\newcommand{\thesisFirstSupervisor}{Henrik Barthels, M.Sc.}

\newcommand{\thesisUniversity}{\protect{RWTH Aachen University}}
\newcommand{\thesisUniversityInstitute}{Aachen Institute for Advanced Study in Computational Engineering Science}
\newcommand{\thesisUniversityGroup}{High-Performance and Automatic Computing Group}
\newcommand{\thesisUniversityCity}{Aachen}
\newcommand{\thesisUniversityStreetAddress}{Rogowski Building\\Schinkelstr. 2}
\newcommand{\thesisUniversityPostalCode}{52062}

\usepackage[utf8]{inputenc}		
\usepackage[english]{babel} 
\usepackage[					
	figuresep=colon,%
	sansserif=false,%
	hangfigurecaption=false,%
	hangsection=true,%
	hangsubsection=true,%
	colorize=full,%
	colortheme=bluemagenta,%
	bibsys=bibtex,%
	bibfile=thesis,%
	bibstyle=numeric,%
]{cleanthesis}

\hypersetup{					
	pdftitle={\thesisTitle},	
	pdfsubject={\thesisSubject},
	pdfauthor={\thesisName},	
	plainpages=false,			
	colorlinks=false,			
	pdfborder={0 0 0},			
	breaklinks=true,			
	bookmarksnumbered=true,		%
	bookmarksopen=true			%
}

\usepackage{layout}
\usepackage{microtype}

\usepackage[T1]{fontenc}
\usepackage{makeidx}
\usepackage[plain,english]{fancyref}
\usepackage{amsmath,amsthm,amssymb,thmtools,centernot}
\usepackage{enumitem}
\usepackage{rotating,adjustbox}
\usepackage{tocloft}
\usepackage{calc}
\usepackage{setspace,nameref,csquotes}
\usepackage{caption}
\usepackage{floatflt,float,wrapfig}
\usepackage[list=on]{subcaption}
\usepackage[nottoc]{tocbibind}
\usepackage{gensymb}
\usepackage[hang]{footmisc}
\usepackage{hyperref}
\usepackage[chapter]{algorithm}
\usepackage[noend]{algpseudocode}
\usepackage{stmaryrd}
\usepackage[acronym,toc,nogroupskip]{glossaries}
\usepackage{booktabs}
\usepackage{mmacells}
\usepackage{tikz}
\usepackage{pgfplots}

\pgfplotsset{compat=1.10}
\usetikzlibrary{calc}
\usetikzlibrary{arrows,shapes,positioning,fit,chains}
\usepgfplotslibrary{fillbetween}

\tikzstyle{st2b}=[rectangle, minimum height=8pt, minimum width=.4pt, inner sep=0pt, draw]

\makeglossaries

\makeatletter
\def\url@leostyle{%
  \@ifundefined{selectfont}{\def\UrlFont{\sf}}{\def\UrlFont{\small\ttfamily}}}
\makeatother{}

\def\Slash{\slash\hspace{0pt}}

\usepackage{xspace}                             
\makeatletter
\DeclareRobustCommand\onedot{\futurelet\@let@token\@onedot}
\def\@onedot{\ifx\@let@token.\else.\null\fi\xspace}
\def\eg{{e.g}\onedot} 
\def\ie{{i.e}\onedot} 
 
\def\etc{{etc}\onedot}

\makeatother

\makeatletter
\declaretheoremstyle[
    spaceabove={.8\baselineskip},
    spacebelow={-.2\baselineskip},
    headfont={\bfseries},
    notefont={\bfseries},
    notebraces={}{},
    bodyfont={\addtolength{\@totalleftmargin}{2em}
       \addtolength{\linewidth}{-2em}
       \parshape 1 2em \linewidth{}
       \itshape},
    headformat={\NAME\ \NUMBER:\NOTE},
    headindent={-1.8em},
    postheadspace={.5em},
    qed={},
    postheadhook={%
      \ifx\@empty\thmt@shortoptarg
        \renewcommand\addcontentsline[3]{}
      \fi}
]{indented}

\declaretheoremstyle[
    spaceabove={.8\baselineskip},
    spacebelow={-.2\baselineskip},
    headformat={\NAME\ \NUMBER \NOTE},
    headpunct=:,
    notebraces={(}{)},
    headfont={\bfseries},
    notefont={\bfseries},
    bodyfont={\itshape}
]{def}

\newlength{\trianglerightwidth}
\algnewcommand{\LineComment}[1]{\Statex \hskip\ALG@thistlm \(\triangleright\) \textit{#1}}
\algnewcommand{\IndentedLineComment}[2]{\Statex \hspace{\dimexpr((\algorithmicindent)*#2)} \(\triangleright\) \textit{#1}}

\algnewcommand{\LineCommentCont}[1]{\Statex%
  \parbox[t]{\dimexpr\linewidth-\ALG@thistlm}{\hangindent=\trianglerightwidth \hangafter=1 \strut$\triangleright$ \textit{#1}\strut}}

\makeatother{}

\algrenewcommand{\alglinenumber}[1]{\color{black}\footnotesize#1\ }

\newcommand{\algorithmicbreak}{\textbf{break}}

\algnewcommand{\IIf}[1]{\State\algorithmicif\ #1\ \algorithmicthen}
\algnewcommand{\EndIIf}{\unskip}

\newcommand{\algrule}[1][.2pt]{\par\vskip.5\baselineskip\hrule height #1\par\vskip.5\baselineskip}

\newcommand*\Let[2]{\State #1 $\gets$ #2}

\newtheoremstyle{definitionstyle}
  {\topsep}
  {\topsep}
  {\em}
  {}
  {\bfseries}
  {.}
  {5pt plus 1pt minus 1pt}
  {}

\declaretheorem[style=definitionstyle,numberwithin=chapter]{definition}

\makeatletter
\def\ll@definition{%
  \protect\numberline{\csname the\thmt@envname\endcsname}%
  \ifx\@empty\thmt@shortoptarg
    \thmt@thmname
  \else
    \thmt@shortoptarg
  \fi}
\def\l@thmt@definition{}
\makeatother

\definecolor{rwthblue}     {RGB}{  0,  84, 159} 
\definecolor{rwthmagenta}  {RGB}{227,   0, 102} 
\definecolor{rwthyellow}   {RGB}{255, 237,   0} 
\definecolor{rwthorange}   {RGB}{246, 168,   0} 
\definecolor{rwthviolet}   {RGB}{ 97,  33,  88} 

\colorlet{o1plotcolor}{rwthmagenta}
\colorlet{m1plotcolor}{rwthblue}
\colorlet{dnplotcolor}{rwthorange}

\cthesissetcolor{RGB}{  0,  84, 159}{227,   0, 102}

\lstdefinestyle{rwthstyle}{
    commentstyle=\color{rwthyellow},
    keywordstyle=\color{rwthblue},
    numberstyle=\tiny\color{black},
    stringstyle=\color{rwthmagenta},
    basicstyle=\footnotesize,
    breakatwhitespace=false,
    breaklines=true,
    captionpos=b,
    keepspaces=true,
    numbers=left,
    numbersep=5pt,
    showspaces=false,
    showstringspaces=false,
    showtabs=false,
    tabsize=2
}

\addto\extrasenglish{%
}

\DeclareMathOperator*{\argmax}{\arg\!\max}

\newcommand{\termset}{\mathcal{T}}
\newcommand{\funcset}{\mathcal{F}}
\newcommand{\varset}{\mathcal{X}}
\newcommand{\groundset}{\mathcal{G}}
\newcommand{\vars}{\mathcal{V}ar}
\newcommand{\pos}{\mathcal{P}os}
\newcommand{\substs}{\mathcal{S}ub}
\newcommand{\fend}{\rangle}
\newcommand{\fstart}{\langle}
\newcommand{\funion}{\sqcup}
\newcommand{\fcompatible}{\triangle}
\newcommand{\multiplicity}{\pi}

\newcommand{\multiset}[1]{\Lbag#1\Rbag}

\newacronym[description={Variadic Syntactic Discrimination Net. See \autoref{def:VSDN}}]{VSDN}{VSDN}{Varidic Syntactic Discrimination Net}
\newacronym[description={Associative Discrimination Net. See \autoref{def:ADN}}]{ADN}{ADN}{Associative Discrimination Net}
\newacronym{MLDN}{MLDN}{Multilayer Discrimination Net}
\newacronym{BLAS}{BLAS}{Basic Linear Algebra Subprograms \citep{Lawson1979,Dongarra1988,Dongarra1990}}
\newacronym{AST}{AST}{Abstract Syntax Tree}
\newacronym[description={Discrimination Net \citep{Bundy1984}}]{DN}{DN}{Discrimination Net}

\makeindex

\begin{document}

\setlength\abovedisplayskip{5pt plus 2pt}
\setlength\belowdisplayskip{5pt plus 2pt}

\pgfplotsset{
    discard if/.style 2 args={
        x filter/.code={
            \edef\tempa{\thisrow{#1}}
            \edef\tempb{#2}
            \ifx\tempa\tempb
                \def\pgfmathresult{inf}
            \fi
        }
    },
    discard if not/.style 2 args={
        x filter/.code={
            \edef\tempa{\thisrow{#1}}
            \edef\tempb{#2}
            \ifx\tempa\tempb
            \else
                \def\pgfmathresult{inf}
            \fi
        }
    },
    every tick label/.append style={font=\small},
    /pgf/number format/1000 sep={}
}

\mmaSet{moredefined={RandomSample,SubsetQ}}

\pagenumbering{roman}			
\pagestyle{empty}				
%
\begin{titlepage}
	\pdfbookmark[0]{Cover}{Cover}
	\flushright
	\hfill
	\vfill
	{\LARGE\thesisTitle \par}
	\rule[5pt]{\textwidth}{.4pt} \par
	{\Large\thesisName}
	\vfill
	\textit{\large\thesisDate}
\end{titlepage}

\begin{titlepage}
	\pdfbookmark[0]{Titlepage}{Titlepage}
	\tgherosfont
	\centering

	{\Large \thesisUniversity} \\[4mm]
	\textsf{\thesisUniversityInstitute} \\
	\textsf{\thesisUniversityGroup} \\

	\vfill
	{\large \thesisSubject} \\[5mm]
	{\LARGE \color{ctcolortitle}\textbf{\thesisTitle} \\[10mm]}
	{\Large \thesisName} \\

	\vfill
	\begin{minipage}[t]{.27\textwidth}
		\raggedleft
		\textit{1. Reviewer}
	\end{minipage}
	\hspace*{15pt}
	\begin{minipage}[t]{.65\textwidth}
		{\Large \thesisFirstReviewer} \\
	  	{\small \thesisFirstReviewerDepartment} \\[-1mm]
		{\small \thesisFirstReviewerUniversity}
	\end{minipage} \\[5mm]
	\begin{minipage}[t]{.27\textwidth}
		\raggedleft
		\textit{2. Reviewer}
	\end{minipage}
	\hspace*{15pt}
	\begin{minipage}[t]{.65\textwidth}
		{\Large \thesisSecondReviewer} \\
	  	{\small \thesisSecondReviewerDepartment} \\[-1mm]
		{\small \thesisSecondReviewerUniversity}
	\end{minipage} \\[10mm]
	\begin{minipage}[t]{.27\textwidth}
		\raggedleft
		\textit{Supervisor}
	\end{minipage}
	\hspace*{15pt}
	\begin{minipage}[t]{.65\textwidth}
		\thesisFirstSupervisor
	\end{minipage} \\[10mm]

	\thesisDate \\

\end{titlepage}

\hfill
\vfill
{
	\small
	\textbf{\thesisName} \\
	\textit{\thesisTitle} \\
	\thesisSubject, \thesisDate \\
	Reviewers: \thesisFirstReviewer\ and \thesisSecondReviewer \\
	Supervisor: \thesisFirstSupervisor \\[1.5em]
	\textbf{\thesisUniversity} \\
	\textit{\thesisUniversityGroup} \\
	\thesisUniversityInstitute \\
	\thesisUniversityStreetAddress \\
	\thesisUniversityPostalCode\ \thesisUniversityCity
}
\cleardoublepage

\setcounter{page}{1}
\pagestyle{plain}				
%
\pdfbookmark[0]{Abstract}{Abstract}
\chapter*{Abstract}
\label{sec:abstract}
\vspace*{-10mm}

Pattern matching is a powerful tool which is part of many functional programming languages as well as computer algebra systems such as Mathematica.
Among the existing systems, Mathematica offers the most expressive pattern matching.
Unfortunately, no open source alternative has comparable pattern matching capabilities.
Notably, these features include support for associative and\Slash{}or commutative function symbols and sequence variables.
While those features have individually been subject of previous research, their comprehensive combination has not yet been investigated.
Furthermore, in many applications, a fixed set of patterns is matched repeatedly against different subjects.
This many-to-one matching can be sped up by exploiting similarities between patterns.
Discrimination nets are the state-of-the-art solution for many-to-one matching.
In this thesis, a generalized discrimination net which supports the full feature set is presented.
All algorithms have been implemented as an open-source library for Python.
In experiments on real world examples, significant speedups of many-to-one over one-to-one matching have been observed.		
\cleardoublepage
%
%
\setcounter{tocdepth}{2}		
\tableofcontents				
\cleardoublepage







\pagenumbering{arabic}			
\setcounter{page}{1}			
\pagestyle{maincontentstyle} 	

\chapter{Introduction}
\label{chp:introduction}

Pattern matching is a powerful tool which is part of many functional programming languages as
well as computer algebra systems such as Mathematica.
It is useful for many applications including symbolic computation, term simplification,
term rewriting systems, automated theorem proving, and model checking.
Term rewriting systems can be used with pattern matching to find matches for the rewrite rules and transform terms.
In functional programming languages, pattern matching enables a more readable and intuitive expression of algorithms.

Among the existing systems, Mathematica offers the most expressive pattern matching.
It is similar to Perl Compatible Regular Expressions \citep{Hazel2017}, but for symbolic tree structures instead of strings.
Patterns are used widely in Mathematica, \eg in function definitions or for manipulating expressions.
Users can define custom function symbols which can also be associative and\Slash{}or commutative.
Mathematica also offers sequence variables which can match a sequence of expressions instead of a single expression.
They are especially useful when working with variadic function symbols.

There is currently no open source alternative to Mathematica with comparable pattern matching capabilities.
In particular, we are interested in similar pattern matching for an experimental linear algebra compiler.
Unfortunately, Mathematica is proprietary and nothing has been published on the underlying pattern matching algorithm.

Previous work predominantly covers syntactic pattern matching, \ie associative\Slash{}commutative\Slash{}variadic
function symbols are not supported. Specifically, no existing work allows function symbols
which are either commutative or associative but not both. However, there are domains where
functions have those properties, \eg matrix multiplication in linear algebra.
Furthermore, most existing publications focus on finding a single match, but we are also interested in finding
all matches for a pattern.

In many applications, a fixed set of patterns will be matched repeatedly against different subjects.
The simultaneous matching of multiple patterns is called many-to-one matching, as opposed to
one-to-one matching which denotes matching with a single pattern.
Many-to-one matching can be sped up by exploiting similarities between patterns.
This has already been the subject of research for both syntactic and AC pattern matching, but not with
the full feature set described above.
Discrimination nets are the state-of-the-art solution for many-to-one matching.
Our goal is to generalize this approach to support the aforementioned full feature set.

We implemented pattern matching with sequence variables and associative\Slash{}commutative function symbols
as an open-source library for Python. In addition, we implemented an efficient many-to-one matching
algorithm that uses generalized discrimination nets. In our experiments we observed significant
speedups of the many-to-one matching over one-to-one matching.

In Chapter~\ref{chp:preliminaries}, we cover the basics of pattern matching as well as our extensions
to syntactic pattern matching. Furthermore, an overview of the related work and existing solutions
is presented. Chapter~\ref{chp:one} introduces algorithms for one-to-one pattern matching which
can handle the previously proposed extensions.
Then, in Chapter~\ref{chp:many}, a generalized version of discrimination nets for many-to-one matching is described.
The performance of those algorithms is evaluated in Chapter~\ref{chp:experiments}.
Finally, conclusions and ideas for future work are presented in Chapter~\ref{chp:conclusions}.
\chapter{Preliminaries} \label{chp:preliminaries}

The notation and definitions used are based on what is used in term rewriting systems literature \citep{Dershowitz1990,Baader1998,Klop2001}. For some extensions to the basic syntactic pattern matching, new notation is introduced.

\section{Syntactic Pattern Matching}

Syntactic pattern matching works on terms. Terms are algebraic data structures constructed from a
countable set of function symbols $\funcset$ and a countable set of variables $\varset$.

The function symbol set is composed of function symbols with different arities, \ie $\funcset = \bigcup_{n\geq0} \funcset_n$ where $\funcset_n$ contains symbols with arity $n$. The function symbols can either have a fixed arity (\ie they only occur in one $\funcset_n$) or be variadic (\ie occur in all $\funcset_n$ for $n \geq n_0$ and some fixed $n_0$). Specifically, $\funcset_0$ contains all constant (function) symbols.

In the following, $f$, $g$, and $h$ are used as function symbols and $a$, $b$, and $c$ refer to constant symbols. Variables are usually called $x$, $y$, and $z$. We usually write common function symbols such as $+$ and $\times$ in infix notation and leave the braces out if unnecessary, \ie we write $+(a, b)$ as $a + b$.

\begin{definition}[Terms]
The set of all terms $\termset(\funcset, \varset)$ is the smallest set such that
\begin{enumerate}
\item $\varset \subseteq \termset(\funcset, \varset)$ and
\item for all $n \geq 0$, all $f \in \funcset_n$, and all $t_1, \dots t_n \in \termset(\funcset, \varset)$ we have $f(t_1, \dots, t_n) \in \termset(\funcset, \varset)$.
\end{enumerate}
\end{definition}
Whenever the actual symbols and variables are not important, the set is shortened to $\termset$.
We also call all terms from $\varset \cup \funcset_0$ \emph{atomic terms} and all others
\emph{compound terms}.  The set $\termset(\funcset, \emptyset) \subseteq \termset(\funcset, \varset)$
is the set of \emph{ground terms} and is called $\groundset(\termset)$ or simply $\groundset$.
Since ground terms are variable-free, they are also called constant. The set of variables occurring
in a term is denoted by $\vars(t)$. A pattern $t$ is called \emph{linear} if every variable in
$\vars(t)$ occurs at most once.

\begin{figure}{h}
\centering
\begin{tikzpicture}[
	grow=down,
	sloped,
	thick,
	every node/.style={rectangle,draw}
]
		\node {$f$}
    		child {
    			node {$g$}
    			child {
    				node {$a$}
            	}
    			child {
    				node {$x$}
            	}
            }
    		child {
    			node {$y$}
    		};
		\end{tikzpicture}
		\caption{Example of a Term as a Tree}
		\label{fig:term_tree}
\end{figure}
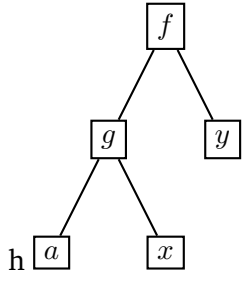

Terms can be viewed as finite ordered trees. Therefore, the pattern matching considered in this
thesis is more closely related to tree pattern matching \citep{Hoffmann1982,Aho1985,Ramesh1988,Steyaert1983}
than to sequence pattern matching \citep{Fernau2013,Fernau2015}.  The leafs of these trees are labeled with
either constants or variables. The inner nodes are labeled with function symbols.
In \autoref{fig:term_tree}, the tree representation of $f(g(a, x), y)$ is displayed.
The \emph{position} of a subterm can be uniquely identified by the position in this tree and can be expressed by a sequence of positive integers,
\eg in the previous example $g$ has the position $1$ and $x$ has the position $1\,2$.
The term itself (the root of the tree) has the empty position sequence $\epsilon$. For a position $\nu$, the subterm in $t$
at that position is denoted by $t|_\nu$, \eg $f(g(a, x), y)|_{1\,2} = x$. The set of all positions
in a term $t$ is called $\pos(t)$, \eg $\pos(f(g(a, x), y)) = \{\epsilon, 1, 1\,1, 1\,2, 2 \}$.
The \emph{size} or \emph{length} of a term $t$ is written as $|t|$ and is given by the number of
nodes in its tree, \ie $|t| = |\pos(t)|$. We also define the \emph{head} of a term to be
the function symbol of a compound term and to be the atomic term itself otherwise.

The interpretation as a tree also leads to the concept of preorder traversal. This sequence is intuitively given by the term without the braces, \ie
for $f(a, g(b, c), d)$ it would be $f\,a\,g\,b\,c\,d$. However, in the following we define the preorder traversal in terms of positions,
\ie for the previous example, we would get the sequence $\epsilon, 1, 2, 2\,1, 2\,2, 3$. Formally, the preorder sequence of a term is defined as

\begin{definition}[Preorder Traversal of Term]
	\label{def:preorder}
	We define a function $next: \termset \times \mathbb{N}^* \rightarrow \mathbb{N}^* \cup \{\top\}$ that gives the next position in the
	preorder traversal sequence (or $\top$ if its end is reached) as \[next(t, \nu) := \min \{\nu' \in \pos(t) \mid \nu < \nu'\}.\]
	where $<$ is the lexicographical order of positions extended with a new minimum
	element $\top$ such that $\forall \nu \in \mathbb{N}^*: \top > \nu$.
\end{definition}


The preorder traversal sequence $pre(t)$ is then given by $\nu_1 < \dots < \nu_k < \top$ where
$\nu_1 = \epsilon$, $\nu_{i+1} = next(t, \nu_i)$ for $i = 1, \dots, k$ and $\nu_{k+1} = \bot$.
Note that $\pos(t) = \{\nu_1, \dots, \nu_k \}$ and $k = |\pos(t)|$.

Similarly, we define the the next sibling's position which is the next in preorder traversal if we skip all children of the current node. For example,
in the term $f(a, g(b, c), d)$, the next sibling of $a$ is $g(b, c)$ and the next sibling of both $g(b, c)$ and $c$ that is $d$.

\begin{definition}[Next Sibling of Term]
	Analogously to \autoref{def:preorder}, we define a function $skip: \termset \times \mathbb{N}^* \rightarrow \mathbb{N}^* \cup \{\top\}$ that gives the position of
	the next sibling if it exists.
	If it does not exists, the first parent that has a next sibling is found and that sibling's position is used. If no such parent exists, $skip$ returns $\top$.
	Formally, $skip$ is defined as \[skip(t, \nu) := \min \{\nu' \in \pos(t) \mid \nu < \nu' \wedge |\nu'| \leq |\nu| \}.\]
\end{definition}

Replacing the subterm at position $\nu$ in term $t$ with some other term $s$ is denoted by $t[s]_\nu$. A special case of replacements are \emph{substitutions}.

\begin{definition}[Substitution]
A substitution is a partial function $\sigma : \varset \nrightarrow \groundset$.
The substitution can be extended to a total function $\hat{\sigma} : \termset(\funcset, \varset) \rightarrow \termset(\funcset, \varset)$ by defining
$\hat{\sigma}(f(t_1, \dots, t_n)) = f(\hat{\sigma}(t_1), \dots, \hat{\sigma}(t_n))$ and $\hat{\sigma}(x) = \sigma(x)$ if $x \in Dom(\sigma)$
or $\hat{\sigma}(x) = x$ otherwise.
\end{definition}

In the following, when we talk about substitutions, we usually mean the extended substitution and for
simplicity we also write $\sigma(t)$ instead of $\hat\sigma(t)$.
We often write substitutions as $\sigma = \{ x_1 \mapsto t_1, \dots, x_n \mapsto t_n \}$ where $\{x_1, \dots, x_n\} = Dom(\sigma)$ are the variables
\emph{instantiated} by the substitution.
The set of all substitutions is called $\substs(\termset(\funcset, \varset))$ or simply $\substs$.
In order to combine multiple substitutions, we need to define which substitutions are compatible:

\begin{definition}[Union of Substitutions]
Two substitutions $\sigma_1$ and $\sigma_2$ are \emph{compatible} (written as $\sigma_1 \fcompatible \sigma_2$) iff for
every $x \in Dom(\sigma_1) \cap Dom(\sigma_2)$ we have that $\sigma_1(x) = \sigma_2(x)$. We write the \emph{union} of two compatible
substitutions as $\sigma_1 \funion \sigma_2$.
\end{definition}

Finally, we can define the core of pattern matching, \ie what constitutes a match:

\begin{definition}[Match]
	A pattern term $t$ \emph{matches} a subject term $s$, iff there exists a substitution $\sigma$ such that $\sigma(t) = s$.
	Such a substitution is also called a \emph{match}.
\end{definition}

The goal of pattern matching is to find a match if it exists. For syntactic pattern matching, this substitution is unique.
In contrast to term unification, the subject must always be constant for pattern matching.

\section{Extensions}
\label{sec:extensions}

Because syntactic pattern matching is very limited in its power, several extensions have been
proposed, \eg to allow matching with associative or commutative function symbols.

\subsection{Sequence Variables}

In order to fully exploit the flexibility of variadic function symbols in patterns, variables with
similar flexibility are needed. Sequence variables are the variable counterpart to variadic
functions. In contrast to regular variables, the sequence variables match a sequence of terms
instead of a single term. Sequence variables are denoted analogously to the notation used in
regular expressions. $x^*$ matches any number of terms including the empty sequence, $x^+$ needs at
least one term to match. Sequence variables have been introduced by Mathematica, but have been
discussed in the context of term rewriting systems as well
\citep{Hamana1997,Kutsia2002,Kutsia2002a,Kutsia2002c}.

Therefore, the definition of a substitution is extended to be a partial function
$\sigma : \varset \nrightarrow \groundset^*$. More precisely, the variable set can be split into
three disjoint subsets: The \emph{regular} variables $\varset_1$, the \emph{star} variables
$\varset_*$ and the \emph{plus} variables $\varset_+$. The substitution is a union
$\sigma = \sigma_1 \funion \sigma_* \funion \sigma_+$ with
$\sigma_1: \varset_1 \rightarrow \groundset$, $\sigma_*: \varset_* \rightarrow \groundset^*$ and
$\sigma_+: \varset_+ \rightarrow \groundset^+$.

When applying a substitution, the replacement of a sequence variable is integrated into the sequence of function symbol arguments: $\sigma(f(a, x^*, b)) = f(a, c, d, b)$ for $\sigma = \{ x^* \mapsto (c, d) \}$.

Patterns with sequence variables may yield multiple valid matches. For instance, the pattern $f(x^+, y^+)$ and the subject $f(a, b, c)$ have
both $\sigma = \{ x^+ \mapsto (a, b), y^+ \mapsto (c) \}$ and $\sigma = \{ x^+ \mapsto (a), y^+ \mapsto (b, c) \}$ as valid matches.

However, pattern matching with sequence variables is much harder than syntactic pattern matching. As an example, consider the pattern $f(x_1^+, \dots, x_m^+)$ and the subject $f(a_1, \dots, a_n)$.
The matches are analogous to integer partitions of $n$ with $m$ parts. There are $\binom{n-1}{m-1}$ many distinct solutions (see Proposition 5.14 of \cite{Gallier2016}), \ie $\mathcal{O}(n^m)$ many.
The problem of finding all matches is therefore exponential in the number of sequence variables.

\subsection{Associative Functions}
\label{sec:assoc_func}

A binary function symbol $f$ is called \emph{associative} iff $f(x, f(y, z)) = f(f(x, y), z)$
for all $x, y, z$. We also write $t_1 =_A t_2$ if two terms are equivalent modulo associativity.
To emphasize that a function symbol is associative, we often write $f_A$ instead of $f$. We also
use $\funcset_A$ to denote the set of all associative function symbols. We call a compound term with an
associative function symbol as head an \emph{associative term}.

In order to easily check for equivalence modulo associativity, we use a \emph{canonical} form of terms and a canonization function $c_A$ such that
$c_A(t_1) = c_A(t_2) \Leftrightarrow t_1 =_A t_2$. Because the order of the operand application does not matter as long as the order of the operands
is unchanged, we use variadic function symbols for associative functions. Then $c_A$ simply \emph{flattens} nested associative terms,
\eg $c_A(f_A(x, f_A(y, z))) = c_A(f_A(f_A(x, y)) = f_A(x, y, z)$. From now on, terms involving associative functions are always assumed to be in canonical form.

In addition, associativity influences what regular variables can match. Consider the pattern
$1 + x$ and the subject $1 + a + b$. Because the latter is equivalent to $1 + (a + b)$, the
substitution $\sigma = \{ x \mapsto (a + b) \}$ is a match. Essentially, regular variables behave
like sequence variables within associative function terms. The problem of associative matching can
be reduced to a matching problem with sequence variables by replacing every variable $x$ which
is an argument of an associative function with a plus variable $x^+$. For example, $f_A(x, a)$
matches a subject iff $f(x^+, a)$ also matches it. Note that the substitution may be different,
\eg for $\sigma_A = \{ x \mapsto f_A(a, b) \}$ and $\sigma_S = \{ x^+ \mapsto (a, b) \}$ we have
that $\sigma_A(f_A(x, a)) = \sigma_S(f(x^+, a)) = f(a, b, a)$.

It has been shown that associative matching is NP-complete \citep{Benanav1987}.

\subsection{Commutative Functions}
\label{sec:comm_func}

A binary function symbol $f$ is called commutative iff $f(x, y) = f(y, x)$ for all $x, y$.
We also write $t_1 =_C t_2$ if two terms are equivalent modulo commutativity.
To emphasize that a function symbol is commutative, we often write $f_C$ instead of $f$. We also
use $\funcset_C$ to denote the set of all commutative function symbols. We call a compound term with a
commutative function symbol as head a \emph{commutative term}.

In order to test equality modulo commutativity, we use a \emph{canonical} form with a
canonization function $c_C$ such that $c_C(t_1) = c_C(t_2) \Leftrightarrow t_1 =_C t_2$.
To define such a canonical form, we require a total order over $\funcset \cup \varset$.
This order can be extended inductively to a total order of terms as defined below.

\begin{definition}[Order of terms]
Given a total order over $\varset \cup \funcset$, the total strict order of terms is inductively extended from it such that $f(t_1, \dots, t_n) < g(s_1, \dots, s_m)$ iff
\begin{enumerate}
\item $f < g$, or
\item $f = g$ and $n < m$, or
\item $f = g$ and $n = m$ and there is an $i$ such that $t_i < s_i$ and for all $1 \leq j < i$ we have $t_j = s_j$.
\end{enumerate}
\end{definition}

The canonical form of a commutative function term is then defined as the minimal term in its
equivalence class, \ie \[c_C(t) := \min \{t' \in \termset \mid t =_C t'\}.\] A term can be
transformed to this form by recursively sorting the arguments of each commutative function subterm.
For example, the canonical form of $f_C(b, a, g(a, b),\allowbreak g(c), h(a))$ is $f_C(a, b, g(c),
g(a, b), h(a))$ assuming $a < b < c < f_C < g < h$. In the following, we assume that any term
containing commutative function symbols is in the commutative canonical form described above.

Alternatively, we can also interpret the arguments of a commutative function as a multiset of terms
instead of a sequence. The interpretation as multiset is also crucial for distributing terms among
sequence variables without enumerating duplicate substitutions. We further extend the codomain of
the substitution to include multiset replacements: $\sigma: \varset \nrightarrow \groundset^* \cup
MS(\groundset)$. We use $\Lbag$ and $\Rbag$ to distinguish multisets from regular
sets, \eg $\multiset{a, a, b} \neq \{a, b\}$. We use the notation $\multiplicity_M(x)$ for the multiplicity of the element $x$ in the
multiset $M$, \eg $\multiplicity_{\multiset{ a, a, b }}(a) = 2$. We use $\uplus$ to denote the
multiset union of sets, \eg $ \{1, 2\} \uplus \{1, 3\} = \multiset{1, 1, 2, 3}$. We write the
scalar multiplication of a multiset as $a \times M$, \ie a repeated union of $M$ $a$
times such that if $M' = a \times M$ we have $\multiplicity_{M'}(x) = a \multiplicity_{M}(x)$ for
all $x$. As an example, $2 \times \multiset{1, 3} = \multiset{1, 1, 3, 3}$.

Commutative matching can yield more valid substitutions than syntactic matching. Consider the pattern $f_C(x, y)$ and the subject $f_C(a, b)$.
This yields two valid matches instead of one, \ie $\sigma_1 = \{ x \mapsto a, y \mapsto b \}$ and $\sigma_2 = \{ x \mapsto b, y \mapsto a \}$.
Generally, for a pattern $f_C(x_1, \dots, x_n)$ and a subject $f_C(a_1, \dots, a_n)$ a total of $n!$ distinct valid matches can be found.
Just finding one substitution in commutative matching has been shown to be NP-complete \citep{Benanav1987}.

When combined with sequence variables, the number of matches can be exponential in the subject size.
Consider the pattern $f_C(x^+, y^+)$ and a subject of the form $f_C(a_1, \dots, a_n)$. Because each
symbol $a_i$ can be assigned to either variable independently, the matches are equivalent to
bitstrings of length $n$. Hence there are $2^n$ distinct matches and enumerating all of them is an exponential problem.

\subsection{Associative-Commutative Functions}

Most function symbols which are either associative or commutative actually possess both properties. Those function symbols are often called AC function symbols and pattern matching with them is called AC pattern matching. We write $t_1 =_{AC} t_2$ if two terms are equivalent modulo associativity and commutativity. Similarly, $f_{AC}$ denotes a function symbol which is both associative and commutative.

The canonical forms for associative and commutative functions can be combined into one form with a canonization function $c_{AC} := c_C \circ c_A$. From now on, every term is assumed to be in this canonical form.

It has been shown that the problem of AC pattern matching is NP-complete \citep{Benanav1987}.
However, this only applies to the problem of finding one match.
If the goal is to find every possible match, the number of matches can be exponential and hence the problem is only NP-hard.
This can easily be seen with an example analogous to the one for commutativity and sequence variables:
For the pattern $f_{AC}(x, y)$ and a subject of the form $f_{AC}(a_1, \dots, a_n)$ there are again $2^n$ distinct matches.

While the subject of AC pattern matching has already been well researched \citep{Hvllot1979,
Benanav1987,Bachmair1993,Eker1995,Kounalis1991,Bachmair1995,Kirchner2001,Lugiez1994}, pattern
matching with just one of the properties has not been explored exhaustively. Still, there are
functions such as matrix multiplication which are not commutative but associative. There are even
functions like the arithmetic mean which are commutative but not associative. Therefore, \eg in
the case of linear algebra, further research is needed to apply pattern matching to domains with
such functions.

\subsection{Guards}

Guards are logical predicates that can be attached to a pattern. This predicate restricts what the pattern can match. The pattern only matches iff the predicate
is satisfied by the match substitution. Hence, a predicate $\varphi$ can be seen as a function $\substs \rightarrow \{ true, false \}$.
We write a guard $\varphi$ on a pattern $t$ as $t\ \mathbf{if}\ \varphi$.

While having one predicate for a whole pattern is easier for theoretical reasoning, practically it is desirable to prune invalid matches early during the match
finding algorithm. Therefore, attaching a predicate to a subterm can be helpful. As an example, we can stop matching on a pattern $f(x, y^+, z^+)\ \mathbf{if}\ x \neq a$
once we discover the first argument of $f(a, b, c, \dots)$ does not satisfy the guard.

\section{Related Work}

There are numerous applications and implementations of pattern matching.
Pattern matching can be used to implement term rewriting systems where it is used to find applicable
rewrite rules. In that context, support for associative and commutative function symbols is often
essential. Pattern matching has also been used as a tool in code generation
for compilers \citep{Aho1985}.

Rewriting can also be used as a programming language \citep{Kirchner2001}. This requires
deterministic pattern matching in order for the programs to be deterministic as well. Especially the
order in which multiple matches are found needs to be deterministic. Such rewrite systems are
Turing complete.

Previous research \citep{Bachmair1993,Christian1993,Bachmair1995} only uses two kinds of
function symbols: syntactic and AC function symbols. The former have fixed arity and no special
properties. The latter are variadic, associative and commutative. They do not support function
symbols which only have some of those properties. Most research assumes patterns to be linear and
only Kutsia has included sequence variables in his research \cite{Kutsia2006,Kutsia2007}.

Even Mathematica can be considered a rewriting language \citep{Buchberger1996}, but it deserves a
special section as is has much more expressive pattern matching than other term rewriting languages.

\subsection{Mathematica} \label{sec:mathematica}

Several programming languages have some form of pattern matching built in. Most of them are
functional or multi-paradigm languages. For example, Haskell \citep{Haskell}, Elixir \citep{Elixir},
Erlang \citep{Erlang}, Rust \citep{Rust}, Scala \citep{Scala}, F\# \citep{FSharp} and Swift \citep{Swift}
all support syntactic pattern matching. Most of those languages also allow guards on
the pattern. The programming language Racket \citep{Racket}
even supports pattern matching with sequence variables, but not commutative or associative
matching. The logic programming language Prolog \citep{Prolog} uses syntactic term unification
which is more general than pattern matching as both terms can contain variables.

Out of all programming languages with pattern matching, the Wolfram Language used in
Mathematica \citep{Mathematica} has the most powerful pattern matching support. It supports both associative
(flat) and commutative (orderless) function symbols as well as sequence variables. Therefore, we
will mostly focus on Mathematica for comparisons.

In the Wolfram Language, variables are written as \texttt{x\_} for $x \in \varset_0$,
\texttt{x\_\_} for $x^+ \in \varset_+$, and \texttt{x\_\_\_} for $x^* \in \varset_*$. Functions
terms are expressed with square brackets instead of braces, \eg \texttt{f[a, b]} instead of
$f(a, b)$. Mathematica also supports anonymous variables (\texttt{\_}, \texttt{\_\_} and
\texttt{\_\_\_}). Guards are written as \texttt{term /; guard}.

Some of Mathematica's features enable more powerful pattern matching than what has been described
in \autoref{sec:extensions}. For example, there is an additional operator
\texttt{p1|p2} to denote alternatives in a pattern, \ie \texttt{p1|p2} matches if either
\texttt{p1} or \texttt{p2} matches. While the alternatives operator can be replicated by using
multiple patterns, their number grows exponentially with the number of alternatives in the worst
case. Mathematica also allows a pattern to be repeated an arbitrary number of times
using the \texttt{Repeated} operation similar to the Kleene plus in regular expressions.
For example, \texttt{f[Repeated[a]]} matches \texttt{f[a, \dots]}, \ie any \texttt{f} function with
one or more \texttt{a} arguments.

Furthermore, similarly to regular expressions, the ``greediness'' of sequence variables can be
controlled, \ie whether they match the shortest or longest possible sequence. In case only one match
is needed, this removes the non-determinism of the matching, because the greediness
determines the order for the exploration of potential matches.
For example, matching the pattern \texttt{f[x\_\_, Longest[y\_\_]]} with the subject
\texttt{f[a, b, c]} will always yield the substitution $\{x^+ \mapsto a, y^+ \mapsto (b, c)\}$ first,
because the $y^+$ variable is made ``greedy''. Therefore, when using the pattern matching for
writing programs, it is easier to reason about its behavior in cases where multiple matches are
possible.

Nonetheless, the Mathematica implementation has some limitations. First, when using
\texttt{ReplaceList} to get all matches for a commutative term, it generates duplicate results:

\begin{mmaCell}[functionlocal=x___]{Code}
SetAttributes[fc, Orderless];
ReplaceList[fc[a, b, a], fc[x___, ___] -> {x}]
\end{mmaCell}
\begin{mmaCell}{Output}
\{\{a,a,b\},\{a,b\},\{a,b\},\{a,a\},\{b\},\{a\},\{a\},\{\}\}
\end{mmaCell}
\noindent
While Mathematica converts commutative terms to a sorted canonical form (\ie \texttt{fc[b, a]}
becomes \texttt{fc[a, b]}), apparently the position of the arguments is used to distinguish
otherwise equivalent arguments when using \texttt{ReplaceList}. By using all argument permutations
instead of just distinct subsets of the argument multiset this duplication is created. The
duplication is amplified when the function symbol is associative as well, since in that
case each of the substitutions in Out[1] is repeated an additional time.

There is also an inconsistency in Mathematica's commutative matching. As expected, commutative
matching works when only using commutative functions:
\begin{mmaCell}{Code}
MatchQ[fc[a, b], fc[b, a]]
\end{mmaCell}
\begin{mmaCell}{Output}
True
\end{mmaCell}
\begin{mmaCell}{Code}
MatchQ[fc[b, a], fc[a, b]]
\end{mmaCell}
\begin{mmaCell}{Output}
True
\end{mmaCell}
\noindent
When using the same variable in both commutative and non-commutative functions,
Mathematica does find the matches in some cases:
\begin{mmaCell}[functionlocal=x___]{Code}
MatchQ[f[f[a, b], fc[a, b]], f[f[x___], fc[x___]]]
\end{mmaCell}
\begin{mmaCell}{Output}
True
\end{mmaCell}
\noindent
But, if the order of the argument in the first subterm is changed, it does not find a match even
though there is one:
\begin{mmaCell}[functionlocal=x___]{Code}
MatchQ[f[f[b, a], fc[a, b]], f[f[x___], fc[x___]]]
\end{mmaCell}
\begin{mmaCell}{Output}
False
\end{mmaCell}
\noindent
This problem even goes so far that the following pattern does not match even though
the order of the arguments is the same in both subterms:
\begin{mmaCell}[functionlocal=x___]{Code}
MatchQ[f[fc[b, a], f[b, a]], f[fc[x___], f[x___]]]
\end{mmaCell}
\begin{mmaCell}{Output}
False
\end{mmaCell}
It seems that this problem arises because Mathematica sorts the arguments of commutative functions
and then does a direct comparision between the sequences if the same variable is encountered later.
For non-commutative functions, the arguments in substitutions are not sorted. This means that if
a sequence variable encounters arguments in a non-commutative subterm which are not sorted, it can
never match them in a commutative subterm, because there the arguments will be sorted. In those
cases, Mathematica will not find a match even though there is one. While this is an edge case
and most patterns will not mix sequence variables in commutative and non-commutative
functions, it is still inconsistent and unexpected behavior. There are workarounds
to get such patterns to behave as expected.

Even though Mathematica has powerful pattern matching features, it has some major drawbacks. The
possibilities to access Mathematica's features from other programming languages is very limited.
Writing large programs in Mathematica can be cumbersome and slow for some applications. Also,
Mathematica is a commercial and proprietary product. Instead, it is desirable to have a free and
open source pattern matching implementation that also enables other researchers to use and extend it.
\chapter{One-to-One Pattern Matching}\label{chp:one}

In this chapter, algorithms for matching various types of patterns are described. First, the
basic case of syntactic patterns is covered. Then the algorithm is extended to support sequence
variables and associative functions. Finally, an algorithm for commutative matching is outlined.

\section{Syntactic Matching}

Syntactic one-to-one pattern matching is straightforward, as both subject and pattern can be
traversed parallelly in preorder and compared at each node. The matching algorithm is given in
pseudocode in Algorithm \ref{alg:syntactic_match}.

\begin{algorithm}[H]
\caption{Syntactic Matching}
\label{alg:syntactic_match}
\begin{algorithmic}[1]
    \Require{\par%
        Subject $s \in \termset$, pattern $p \in \termset$, and an initial substitution $\sigma \in \substs$.
    }
    \Ensure{\par%
        The match $\sigma$ or $\lightning$ iff there is no match.
    }
    \algrule
    \Function{SyntacticMatch}{$s, p, \sigma$}
        \If{$p \in \varset$}
            \Return{$\{ p \mapsto s \}$}
        \EndIf
        \If{$p = f(p_1, \dots, p_n)$ and $s = f(s_1, \dots, s_n)$ for some $f \in \funcset, n \in \mathbb{N}$}
            \ForAll{$i = 1 \dots n$}
                \Let{$\sigma'$}{$\textsc{SyntacticMatch}(s_i, p_i, \sigma)$}
                \If{$\sigma' = \lightning$ or $\sigma' \centernot\fcompatible \sigma$}
                    \Return{$\lightning$}
                \EndIf
                \Let{$\sigma$}{$\sigma \funion \sigma'$}
            \EndFor
            \State\Return{$\sigma$}
        \EndIf
        \State\Return{$\lightning$}
    \EndFunction
\end{algorithmic}
\end{algorithm}
\noindent
For a given subject $s$ and a pattern $p$, $\textsc{SyntacticMatch}(s, p, \emptyset)$ yields the
match $\sigma$ or $\lightning$ if there is no match. Note that there is at most one match
in syntactic matching.
This algorithm has a worst case complexity of $\mathcal{O}(min(|p|, |s|))$. One optimization in this
algorithm is the early abort when subject and pattern have different number of arguments. Because
syntactic matching does not involve associative function symbols or sequence variables, in those
cases there can be no match.

\section{Matching with Sequence Variables}

In Algorithm \ref{alg:match_one_to_one}, the pseudocode for an algorithm that returns all matches
for a given subject and a given non-commutative pattern is displayed. An algorithm for matching
commutative patterns is discussed in the next section.

\begin{algorithm}[H]
\caption{Matching with Sequence Variables}
\label{alg:match_one_to_one}
\begin{algorithmic}[1]
    \Require{\par%
        Subject sequence $s_1 \dots s_n \in \termset$, pattern $p \in \termset$, an optional associative
        function symbol $f_a \in \funcset_A \cup \{ \bot \}$, and initial substitutions $\Theta \subseteq \substs$.
    }
    \Ensure{\par%
        The set of matches.
    }
    \algrule
    \Function{MatchOneToOne}{$s_1 \dots s_n, p, f_a, \Theta$}
        \If{$p \in \funcset_0$} \Comment{Constant symbol pattern}
            \IIf{$n = 1$ and $s_1$ = p}
                \Return{$\Theta$}
            \EndIIf
        \ElsIf{$p \in \varset_0$ and $f_a = \bot$} \Comment{Regular variable pattern}
            \Let{$\sigma'$}{$\{ p \mapsto s_1 \}$}
            \IIf{$n = 1$}
                \Return{$\{\sigma \funion \sigma' \mid \sigma \in \Theta \wedge \sigma \fcompatible \sigma' \}$}
            \EndIIf
        \ElsIf{$p \in \varset$} \Comment{Sequence variable pattern}
            \If{$p \in \varset_0$ and $f_a \neq \bot$} \Comment{Regular variable in assoc. function?}
                \Let{$\sigma'$}{$\{ p \mapsto f_a(s_1, \dots, s_n) \}$}
            \Else
                \Let{$\sigma'$}{$\{ p \mapsto (s_1, \dots, s_n) \}$}
            \EndIf
            \IIf{$p \in \varset_*$ or $n \geq 1$}
                \Return{$\{\sigma \funion \sigma' \mid \sigma \in \Theta \wedge \sigma \fcompatible \sigma' \}$}
            \EndIIf
        \ElsIf{$n = 1$} \Comment{Compound pattern}
            \Let{$h$}{$head(p)$}
            \If{$h = head(s_1)$}
                \Let{$p_1 \dots p_k$}{$args(p)$}
                \Let{$q_1 \dots q_l$}{$args(s_1)$}
                \If{$h \in \funcset_a$}
                    \Let{$f_a'$}{$h$}
                \Else
                    \Let{$f_a'$}{$\bot$}
                \EndIf
                \State\Return{$\textsc{MatchSequence}(q_1 \dots q_l, p_1 \dots p_k, f_a', \Theta)$} \label{alg:match_one_to_one:branch_line}
            \EndIf
        \EndIf
        \State\Return{$\emptyset$}
    \EndFunction
\end{algorithmic}
\end{algorithm}
\noindent
The algorithm matches a sequence of subject terms against the pattern. Using a sequence of subjects
instead of a single subject argument is needed to correctly match sequence variables. The set of
matches for subject $s$ and pattern $p$ is returned by
$\textsc{MatchOneToOne}(s, p, \bot, \{\emptyset\})$.

We use a set of substitutions here, because for
non-syntactic patterns there can be multiple matches. Hence, in Algorithm \ref{alg:match_one_to_one},
$\emptyset$ as a result is equivalent to $\lightning$ in Algorithm \ref{alg:syntactic_match}. Note
that this is different from $\{\emptyset\}$ as a result. The latter means there is a single match
which is the empty substitution, \ie when the pattern does not contain any variables.

We pass an initial set of substitutions into the function to be able to chain them together for
matching the argument subpatterns of a compound pattern. Consider a pattern $f(g(x), g(x, a, b), \dots)$
and a subject $f(g(a), g(c, a, b), \dots)$. There is no match as the substitutions of the subpatterns for
$x$ are not compatible. When calls to \textsc{MatchOneToOne} are chained and the current set of
substitutions gets passed around, matching can be aborted early when there is a mismatch. In the
previous example, at some point $\textsc{MatchOneToOne}(c, x, \bot, \{ \{ x \mapsto a \} \})$ gets
called and returns an empty substitution set, aborting the matching.

An alternative solution would be to match each pattern
argument separately and only check whether the substitutions are compatible once all submatching
has completed. The disadvantage of this approach is that in case of a mismatch, backtracking cannot
occur early. However, by making the submatches independent, they could be parallelized for better
performance. We chose the chained solution because in our application the overhead of parallelization
would likely outweigh the benefits.

The base cases of constant symbol and variable patterns are handled in the \textsc{Match\-OneToOne} function
directly, while the handling of compound terms is deferred to the \textsc{MatchSequence} function
displayed in Algorithm \ref{alg:match_function}. In the case of a variable pattern, the
\textsc{MatchOneToOne} function checks for all substitutions $\sigma$ in the initial substitution
set if the new variable substitution $\sigma'$ is compatible with $\sigma$. For those that are
compatible, the union substitution is constructed and the set of all these substitutions is returned.
While the constant symbols are only a special case of function symbols and would be handled
correctly by \textsc{MatchSequence}, they are treated separately for performance reasons.

\paragraph{Description of Algorithm \ref{alg:match_function}}
In the following, a line-by-line description of the algorithm for sequence matching is given:
\begin{description}[leftmargin=!,labelwidth=\widthof{\bfseries 22},itemsep=-5pt]
\item [2+4] The numbers of star and plus variables in the pattern are counted\footnote{These values
can be cached if the pattern becomes large enough for their computation to become a bottleneck.}, respectively.
\item [3] The matching aborts early, if the total number of arguments required in the pattern
exceeds the number of arguments in the subject. As an example, consider the pattern
$f(x^+, y^*, a, b)$ which requires a subject argument each for $a$ and $b$ and at least one for
$x^+$  to match. Hence, in total, a match requires a subject term with at least 3 arguments. Given a subject
$f(a, b)$, we have that $m - n_* = 4 - 1 = 3 > 2 = n$ and hence the pattern cannot match.
\item [5--6] Regular variables in associative functions are treated as plus sequence variables and
hence their number is added to the plus variable count.
\end{description}
\begin{algorithm}[H]
\begin{algorithmic}[1]
    \Require{\par%
        Subject sequence $s_1 \dots s_n \in \termset$, pattern sequence $p_1 \dots p_m \in \termset$, an optional associative
        function symbol $f_a \in \funcset_A \cup \{ \bot \}$, and initial substitutions $\Theta \subseteq \substs$.
    }
    \Ensure{\par%
        The set of matches.
    }
    \algrule
    \Function{MatchSequence}{$s_1 \dots s_n, p_1 \dots p_m, f_a, \Theta$}
        \Let{$n_*$}{$\sum_{i=1}^{m} [p_i \in \varset_*]$} \Comment{Number of star variables}
        \IIf{$m - n_* > n$}
            \Return{$\emptyset$}
        \EndIIf
        \Let{$n_+$}{$\sum_{i=1}^{m} [p_i \in \varset_+]$} \Comment{Number of plus variables}
        \If{$f_a \neq \bot$}
            \IndentedLineComment{Count regular vars as plus vars in assoc. function}{2}
            \Let{$n_+$}{$n_+ + \sum_{i=1}^{m} [p_i \in \varset_0]$}
        \EndIf
        \Let{$n_{free}$}{$n - m + n_*$} \Comment{Number of free arguments in the subject}
        \Let{$n_{seq}$}{$n_* + n_+$} \Comment{Total number of sequence variables}
        \Let{$\Theta_R$}{$\emptyset$} \Comment{Result substitutions}
        \LineComment{For every distribution of free arguments among the seq. vars...}
        \ForAll{$(k_1, \dots, k_{n_{seq}}) \in \{ v \in \mathbb{N}^{n_{seq}} \mid \sum v = n_{free} \}$}
            \Let{$i$}{$0$} \Comment{Subject argument index}
            \Let{$j$}{$0$} \Comment{Sequence var index}
            \Let{$\Theta'$}{$\Theta$} \Comment{Intermediate substitutions}
            \LineComment{For every pattern argument...}
            \ForAll{$l = 1 \dots m$}
                \Let{$l_{sub}$}{$1$} \Comment{Length of subject argument subsequence}
                \IndentedLineComment{If the argument is a sequence variable...}{3}
                \If{$p_l \in \varset_+ \cup \varset_*$ or $p_l \in \varset_0$ and $f_a \neq \bot$}
                    \Let{$l_{sub}$}{$l_{sub} + k_j$}
                    \If{$p_l \in \varset_*$}
                        \Let{$l_{sub}$}{$l_{sub} - 1$}
                    \EndIf
                    \Let{$j$}{$j + 1$}
                \EndIf
                \Let{$s'$}{$s_i \dots s_{i + l_{sub}}$} \Comment{Subject argument subsequence}
                \Let{$\Theta'$}{$\textsc{MatchOneToOne}(s', p_l, f_a, \Theta')$}
                \IIf{$\Theta' = \emptyset$}
                    \algorithmicbreak \Comment{No match for distribution}
                \EndIIf
                \Let{$i$}{$i + l_{sub}$}
            \EndFor
            \Let{$\Theta_R$}{$\Theta_R \cup \Theta'$}
        \EndFor
        \State\Return{$\Theta_R$}
    \EndFunction
\end{algorithmic}
\caption{Matching a non-commutative function}
\label{alg:match_function}
\end{algorithm}
\begin{description}[leftmargin=!,labelwidth=\widthof{\bfseries 22},itemsep=-5pt]
\item [7] The total number of ``free'' arguments in the subject can be determined, \ie how many
arguments need to be distributed among the sequence variables. As an example, consider the pattern
$f(x^+, y^*, a, b)$ again, but this time with a subject $f(a, b, c, a, b)$. We have $n_{free} =
5 - 4 + 1 = 2$ free arguments (in this case $b$ and $c$ can go in either $x^+$ or $y^*$).
\item [8] The number of sequence variables is needed to enumerate all distributions of the free
arguments among them.
\item [9] We use result substitution set $\Theta_R$ to collect all the intermediate matches
found for each generated distribution (see line 25).
\item [10] Here we enumerate all possible distributions of the arguments among the sequence
variables. This is related to the combinatorial problem of enumerating the
$n_{free}$-multicombinations of $n_{seq}$ objects\footnote{This is case four of the Twelvefold Way
\citep{Stanley2011}} \cite{Stanley2011,Knuth2005}. The solutions are also known
as the weak compositions of $n_{free}$ with $n_{seq}$ parts. Algorithms to enumerate all these
solutions efficiently have been developed, \eg by \cite{Ruskey2009} and \cite{Page2012}. Generating the
next weak composition can be done in $\mathcal{O}(1)$, but there are $\binom{n-1}{m-1}$ many
distinct weak compositions \citep{Gallier2016} resulting in a total complexity of $\mathcal{O}(n^m)$.
Note that the set of weak compositions for $n_{seq} = n_{free} = 0$ contains the empty tuple $()$ as
single argument, so the case without sequence variables is also covered. Similarly, if
$n_{seq} = 0$ and $n_{free} \geq 1$, the composition set is empty and there is no match.
\item [11--21] For each weak composition $k$ we map every pattern argument $p_l$ to a subsequence
$s'$ of the subject arguments $s$. As an example, consider the pattern $f(x^+, y^+)$ and the subject
$f(a, b, c)$. We enumerate weak compositions of $1$ with $2$ parts, \ie $k \in \{ (0, 1), (1, 0) \}$.
These correspond to the substitutions $\{ x^+ \mapsto (a), y^+ \mapsto (b, c) \}$ and
$\{ x^+ \mapsto (a, b), y^+ \mapsto (c) \}$ which are both collected and returned as the overall result.
\item [11] The index $i$ is used to track the next subject argument to assign.
\item [12] The index $j$ is used to determine which part of the composition $k$ is used for the next
sequence variable.
\item [13] An intermediate substitution set $\Theta'$ is used to iteratively merge submatches for a
single composition.
\item [14] We use the index $l$ to loop over all pattern arguments $p$.
\item [15--21] Depending on whether the pattern argument is a sequence variable or not, we assign a
subsequence of subject arguments $s'$ to the pattern argument $p_l$. The subsequence always starts
at index $i$ and has a length of $l_{sub}$. We start off with a default length of $1$, \ie one
required subject argument for every pattern argument. For all sequence variables we add the
corresponding part of the composition $k_j$ to the length, and for star variables we subtract
the initial $1$ because the star variable does not require an argument.
\item [22] Here the recursion occurs as \textsc{MatchOneToOne} is called to match the subject
argument subsequence against the pattern argument. The intermediate substitution set $\Theta'$
is updated to reflect the effects of this submatch on the substitutions.
\item [23] If $\Theta'$ becomes empty, there is a mismatch for the current sequence argument
distribution. Therefore, the matching can continue with the next composition.
\item [24] Otherwise, we move the subject argument index $i$ to the end of the last subsequence and
continue with the next pattern argument.
\item [25] The matches for every possible argument distribution are collected.
\end{description}
\noindent
Note that the above algorithm does not include the support for commutative patterns. The algorithm
for commutative matching will be discussed in the next section.

\section{Commutative Matching}

For commutative patterns, the argument order does not matter for matching. This can
be achieved with a brute force approach by applying non-commutative matching to all permutations of the
subject arguments and merging the results. However, for a commutative term with $n$ arguments there are
$n!$ permutations. As discussed in the previous section, there are exponentially many potential
matches for each of those permutations. Furthermore, this approach will result in enumerating
equivalent matches multiple times: Let the pattern $f_c(x^+, y^+)$ and the subject $f_c(a, b, c)$ serve
as an example. Using non-commutative matching on the permutations $f_c(a, b, c)$ and $f_c(b, a, c)$ yields
the substitution $\{ x^+ \mapsto \multiset{a, b}, y^+ \mapsto \multiset{c} \}$ in both
cases. For this example, every match is enumerated twice by the brute force approach.

Still, our approach is to take all distinct mappings between patterns and subjects and recursively apply
\textsc{MatchOneToOne} to find matches for each individual pattern\hyphen{}subjects pair. However, the goal
is to only check each distinct mapping once and to overall reduce the search space.
Because the problem of commutative matching is NP-complete,
in the worst case, matching can be exponential in the size of the subject. However, by choosing
an efficient order for the matching, large portions of the match search space can potentially be
pruned. This reduces the cost of matching in most common cases.
With this idea in mind, we propose the following order of matching for argument patterns $p$ of commutative
compound patterns:

\begin{enumerate}
\item Constant patterns, \ie patterns that are ground terms ($p \in \groundset$)
\item Matched variables, \ie variables that already have a value assigned in the current substitution,
    ($p \in \varset$ and $p \in Dom(\sigma)$)
\item Non-variable patterns ($p \in \termset \setminus \varset$)
\item Regular variables ($p \in \varset_0$)
\item Sequence variables ($p \in \varset_* \cup \varset_+$)
\end{enumerate}

Let us take a look again at the size of the search space.
For a subject with $n$ arguments and a pattern with $m$ arguments we have up to
$m^n$ potential combinations\footnote{If all pattern arguments are star variables, we can assign
each subject to any of them independently. If all subject arguments are distinct, that yields a
total of $m^n$ combinations which are all valid matches, assuming there are no further constraints.}
that need to be checked. Reducing both $n$ and $m$ can dramatically
reduce the number of possibilities which need to be considered. As an example, covering constant
patterns early means the corresponding subjects do not need to be considered later for sequence
variables. The overall goal of this approach is to perform cheap submatches early to reduce the search
space for the more expensive matches.

To integrate the commutative matching into Algorithm \ref{alg:match_one_to_one}, it suffices to
replace the line \ref{alg:match_one_to_one:branch_line} with a branch that, depending on whether the function is commutative or not,
proceeds to call the commutative matching algorithm or \textsc{MatchSequence}, respectively. The
commutative matching algorithm is not given as pseudocode because it would be too long, but
its steps are outlined in the next section.

\subsection{Matching Steps}

For matching individual pattern arguments and subject sequences in commutative terms,
\textsc{MatchOneToOne} can be used again. As stated earlier, we want to enumerate
all possible mappings between subject and pattern arguments and match them. In the following, the
matching steps are described and we use $S$ to denote the multiset of subject arguments and $P$ for the multiset of
pattern arguments:

\paragraph{Constant patterns} To match constant arguments, we can construct the multiset
$P_{const} = P \cap \groundset$ of constant pattern arguments. If $P_{const} \not\subseteq S$, a
match is impossible, because at least one constant argument in the pattern has no matching subject argument, and
the matching can be aborted. Otherwise we remove all matched constants from $S$ and $P$ and proceed, \ie
$S \leftarrow S \setminus P_{const}$ and $P \leftarrow P \setminus P_{const}$ .

\paragraph{Matched variables} Similarly to constant patterns, checking variables which already
have a value in the current substitution is very cheap. If $x \in Dom(\sigma)$, we
construct a multiset $P_x := \multiplicity_{P}(x) \times \multiset{\sigma(x)}$ from the existing
substitution. This accounts for a variable occurring multiple times in the pattern by multiplying
the substitution by the number of occurrences, \eg if $\sigma(x) = (a, b)$ and $x$ occurs two times
in the pattern, then $P_x = \multiset{a, a, b, b}$. Again, if $P_x \not\subseteq S$, a match is impossible.
Otherwise we remove the matched terms from $S$ and $x$ from the pattern set, \ie $S \leftarrow S - x$
and $\multiplicity_{P}(x) \leftarrow 0$. This can reduce the search space considerably, because for
those variables no potential matches need to be enumerated. If there is more than one initial
substitution, this has to be done once per initial substitution and the remaining patterns need to
be matched separately.

\paragraph{Non-variable patterns} This set of patterns encompasses all the remaining pattern
arguments except single variables. However, these patterns do contain variables (otherwise they
would be constant) and hence any match for them is not empty. Therefore, exhaustive matching has to
be performed to find these matches. By grouping patterns and subjects by their head, we can limit
the pairwise combinations between subjects and patterns to consider, because only those with the
same head can match. Each pairing
can recursively be matched by using \textsc{MatchOneToOne}. The matches are not necessarily
independent, as two patterns can contain the same variable. Consequently, we use an iterative
process similarly to the process in Algorithm \ref{alg:match_function}. Instead of processing
subpattern matching independently, the calls to \textsc{MatchOneToOne} are chained together
for all combinations of pairings to find the overall match. Note that after matching the
non-variable patterns, the step of matching already matched variables can be repeated, because new
variables might have been added to the substitution during this step.

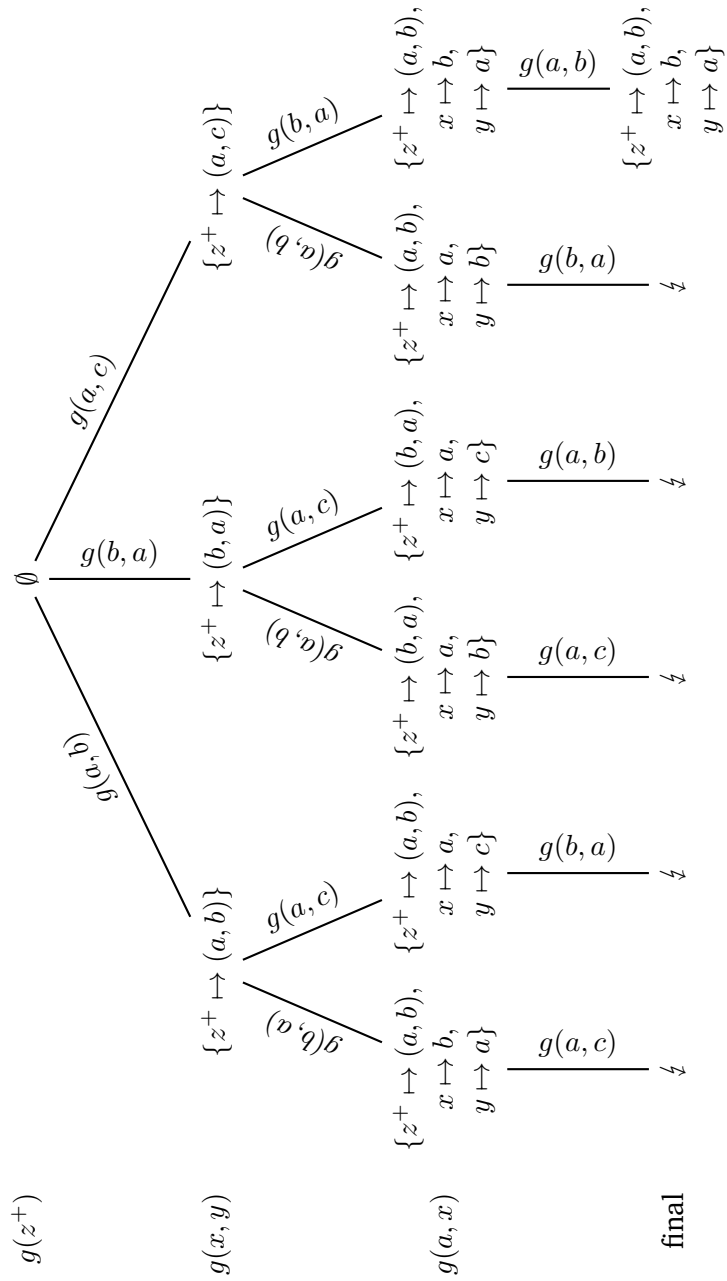
\begin{figure}[H]
    \centering
    \begin{tikzpicture}[sloped,thick,level 1/.style={sibling distance=2cm,level distance=2.5cm}]
        \node (Root) {$\emptyset$}
            child {
                node {$\{ x \mapsto b \}$}
                child {
                    node {$\{ x \mapsto b, y \mapsto a \}$}
                    child {
                        node[align=center] {$\{ x \mapsto b, y \mapsto a,$\\$z^+ \mapsto (a, c) \}$}
                        edge from parent
                        node[above] {$g(a, c)$}
                    }
                    edge from parent
                    node[above] {$g(b, a)$}
                }
                child {
                    node {$\lightning$}
                    edge from parent
                    node[above] {$g(a, c)$}
                }
                edge from parent
                node[above] {$g(a, b)$}
            }
            child {
                node {$\lightning$}
                edge from parent
                node[above] {$g(b, a)$}
            }
            child {
                node {$\{ x \mapsto c \}$}
                child {
                    node {$\lightning$}
                    edge from parent
                    node[above] {$g(a, b)$}
                }
                child {
                    node {$\lightning$}
                    edge from parent
                    node[above] {$g(b, a)$}
                }
                edge from parent
                node[above] {$g(a, c)$}
            };

        \begin{scope}[every node/.style={right}]
            \path (Root         -| Root-1-1-1) ++(-1.5cm,0) node[left] {$g(a, x)$};
            \path (Root-1       -| Root-1-1-1) ++(-1.5cm,0) node[left] {$g(x, y)$};
            \path (Root-1-1     -| Root-1-1-1) ++(-1.5cm,0) node[left] {$g(z^+)$};
            \path (Root-1-1-1   -| Root-1-1-1) ++(-1.5cm,0) node[left] {final};
        \end{scope}
    \end{tikzpicture}
    \caption{Search Tree for order $g(a, x)$, $g(x, y)$, $g(z^+)$}
    \label{fig:match_tree_good}
\end{figure}

Let the pattern $f_c(g(a, x), g(x, y), g(z^+)$ and the subject $f_c(g(a, b), g(b, a), g(a, c))$
serve as an example. There is only one match in this example, namely $\{ x \mapsto b, y \mapsto a,
z^+ \mapsto (a, c) \}$. However, there is a total of $3! = 6$ distinct mappings between the pattern
arguments and the subject arguments all of which could potentially match. In \autoref{fig:match_tree_good}
and \autoref{fig:match_tree_bad} two different search trees for the matching are displayed.
Each level in the tree corresponds to a pattern argument that needs to be matched. Each edge is
labeled with the subject that is mapped to the respective pattern argument at that level. Every node
corresponds to a call to \textsc{MatchOneToOne} and the resulting substitution. Failing matches
are denoted with $\lightning$.

\begin{figure}[h]
\begin{adjustbox}{addcode={\begin{minipage}{\width}}{\caption{%
      Search Tree for order $g(z^+)$, $g(x, y)$, $g(a, x)$
      }\label{fig:match_tree_bad}\end{minipage}},rotate=90,center}
    \begin{tikzpicture}[
        sloped,
        thick,
        every node/.style={transform shape},
        level 1/.style={sibling distance=5.2cm,level distance=2.5cm},
        level 2/.style={sibling distance=2.6cm,level distance=3cm}
    ]
        \node (Root) {$\emptyset$}
            child {
                node {$\{ z^+ \mapsto (a, b) \}$}
                child {
                    node[align=center] {$\{ z^+ \mapsto (a, b),$\\$x \mapsto b,$\\$y \mapsto a \}$}
                    child {
                        node {$\lightning$}
                        edge from parent
                        node[above] {$g(a, c)$}
                    }
                    edge from parent
                    node[above] {$g(b, a)$}
                }
                child {
                    node[align=center] {$\{ z^+ \mapsto (a, b),$\\$x \mapsto a,$\\$y \mapsto c \}$}
                    child {
                        node {$\lightning$}
                        edge from parent
                        node[above] {$g(b, a)$}
                    }
                    edge from parent
                    node[above] {$g(a, c)$}
                }
                edge from parent
                node[above] {$g(a, b)$}
            }
            child {
                node {$\{ z^+ \mapsto (b, a) \}$}
                child {
                    node[align=center]  {$\{ z^+ \mapsto (b, a),$\\$x \mapsto a,$\\$y \mapsto b \}$}
                    child {
                        node {$\lightning$}
                        edge from parent
                        node[above] {$g(a, c)$}
                    }
                    edge from parent
                    node[above] {$g(a, b)$}
                }
                child {
                    node[align=center]  {$\{ z^+ \mapsto (b, a),$\\$x \mapsto a,$\\$y \mapsto c \}$}
                    child {
                        node {$\lightning$}
                        edge from parent
                        node[above] {$g(a, b)$}
                    }
                    edge from parent
                    node[above] {$g(a, c)$}
                }
                edge from parent
                node[above] {$g(b, a)$}
            }
            child {
                node {$\{ z^+ \mapsto (a, c) \}$}
                child {
                    node[align=center] {$\{ z^+ \mapsto (a, b),$\\$x \mapsto a,$\\$y \mapsto b \}$}
                    child {
                        node {$\lightning$}
                        edge from parent
                        node[above] {$g(b, a)$}
                    }
                    edge from parent
                    node[above] {$g(a, b)$}
                }
                child {
                    node[align=center] {$\{ z^+ \mapsto (a, b),$\\$x \mapsto b,$\\$y \mapsto a \}$}
                    child {
                        node[align=center] {$\{ z^+ \mapsto (a, b),$\\$x \mapsto b,$\\$y \mapsto a \}$}
                        edge from parent
                        node[above] {$g(a, b)$}
                    }
                    edge from parent
                    node[above] {$g(b, a)$}
                }
                edge from parent
                node[above] {$g(a, c)$}
            };

        \begin{scope}[every node/.style={right}]
            \path (Root         -| Root-1-1-1) ++(-1.5cm,0) node[left] {$g(z^+)$};
            \path (Root-1       -| Root-1-1-1) ++(-1.5cm,0) node[left] {$g(x, y)$};
            \path (Root-1-1     -| Root-1-1-1) ++(-1.5cm,0) node[left] {$g(a, x)$};
            \path (Root-1-1-1   -| Root-1-1-1) ++(-1.5cm,0) node[left] {final};
        \end{scope}
    \end{tikzpicture}
\end{adjustbox}
\end{figure}
\clearpage
In \autoref{fig:match_tree_good}, the matching can be aborted early in a lot of cases because of a
mismatch between variable substitutions. Therefore, only a total of 8 child nodes are visited,
corresponding to 8 calls to \textsc{MatchOneToOne}. For the order in \autoref{fig:match_tree_bad},
the match always fails only at the last pattern argument. Hence, the matching cannot abort early
and a total of 15 calls to \textsc{MatchOneToOne} need to be made. This example illustrates,
how iterative matching can lower the costs of matching in some cases, because matching can abort
early without fully matching each subpattern. If independent matching of the subpatterns was
employed, the full search space as depicted in \autoref{fig:match_tree_bad} needs to be searched.
However, we do not propose to order the non-variable pattern arguments in a special way,
we just take advantage of the order if it already exists. In the worst case, the whole search space
still has to be explored.

\paragraph{Regular variables} For every regular variable $x \in \varset_0$, if the the surrounding
function is not associative, every mapping of subject arguments onto the variable needs to be checked.
The multiplicities of the variable and of each subject argument needs to be taken into account, \ie
only arguments from $S_x := \{ s \in S \mid \multiplicity_{S}(s) \geq \multiplicity_{P}(x) \}$ are used to match the variable $x$.
For every substitution that is checked, the corresponding subject arguments are removed from the
subject set temporarily, \ie we set $\multiplicity_{S}(s) \leftarrow \multiplicity_{S}(s) - \multiplicity_{P}(x)$.
This process is applied iteratively again for every regular variable shrinking the subject set
at every step. Each step also involves backtracking to the previous step after completing the match
to choose a different variable substitution for $x$ from $S_x$. Again, each regular variable matching
step is depending on the previous one and needs to be repeated for every choice made in the previous step.

To illustrate the issue with multiplicities, examine the pattern $f_c(x, x, y^*)$ and the subject
$f_c(a, a, a, b, b, c)$. Only $a$ and $b$ are candidates for the variable $x$, since there is only
one $c$ but two occurrences of $x$. Hence we have $S_x = \{ a, b \}$ in this case. Assuming that
$a$ has been chosen as a substitution for $x$, we get an updated subject set $\multiset{a, b, b, c}$
and proceed with matching $y^*$. After that is completed, we backtrack and choose $b$ as a
substitution obtaining the remaining subject set $\multiset{a, a, a, c}$. After processing that, the matching
is finally done because there are no more choices left for $x$.

\paragraph{Sequence Variables} Sequence variables are the most expensive part of matching, as there
can be exponentially many combinations between subject arguments and sequence variables. In general,
there can be up to $n^m$ different matches given $n$ sequence variables and $m$ subject arguments.
The algorithm for finding all sequence variable matches is described in the
next section.

\subsection{Sequence Variables}
\label{sec:seq_vars}

To enumerate all distributions of subjects onto sequence variables without duplicate results,
the problem is modeled with several disjoint linear diophantine equations. For every sequence variable $x$ and every
symbol $a$, we have a variable $x_a$ in the equations that determines how many $a$ are contained
in the solution multiset for $x$. Hence, each non-negative integer solution to the equation system
corresponds to a valid substitution for the sequence variables.

As an example consider the multiset of remaining terms $\multiset{ a, b, b, c, c, c }$ and the
variable multiset $\multiset{x^*, y^+, y^+}$. This yields the following equations:
\begin{align*}
    1 &= x_a + 2 y_a \\
    2 &= x_b + 2 y_b \\
    3 &= x_c + 2 y_c
\end{align*}
We are looking for all non-negative solutions that satisfy the additional constraint
$y_a + y_b + y_c \geq 1$ (because $y^+$ is a plus variable).
Note that while these equations have infinitely many integer solutions, there are only finitely many
non-negative solutions. For the above example there are three solutions:
\begin{alignat*}{11}
    x_a &= 1 &\ \wedge\ && x_b &= 2 &\ \wedge\ && x_c &= 1 &\ \wedge\ && y_a &= 0 &\ \wedge\ && y_b &= 0 &\ \wedge\ && y_c &= 1 \\
    x_a &= 1 &\ \wedge\ && x_b &= 0 &\ \wedge\ && x_c &= 3 &\ \wedge\ && y_a &= 0 &\ \wedge\ && y_b &= 1 &\ \wedge\ && y_c &= 0 \\
    x_a &= 1 &\ \wedge\ && x_b &= 0 &\ \wedge\ && x_c &= 1 &\ \wedge\ && y_a &= 0 &\ \wedge\ && y_b &= 1 &\ \wedge\ && y_c &= 1
\end{alignat*}
These correspond to the substitutions
\begin{alignat*}{2}
    \sigma_1 &= \{ x \mapsto \multiset{ a, b, b, c },   &\quad&y \mapsto \multiset{ c } \} \\
    \sigma_2 &= \{ x \mapsto \multiset{ a, c, c, c },   &\quad&y \mapsto \multiset{ b } \} \\
    \sigma_3 &= \{ x \mapsto \multiset{ a, c },  	    &\quad&y \mapsto \multiset{ b, c } \}.
\end{alignat*}
Formally we define the set of equations as follows:
\begin{definition}[Sequence Variable Equations]
Let $S$ and $V$ be the multisets of subject terms and variables respectively, let $s_1, \dots, s_n \in \termset$
and $x_1, \dots, x_m \in \varset_* \cup \varset_+$ be the distinct elements of $S$ and $V$ respectively, let
$d_i := \multiplicity_S(s_i)$ for $i = 1,\dots,n$, and let $c_j := \multiplicity_V(x_j)$ for $j = 1,\dots,m$.
Then the sequence variable equations for $S$ and $V$ are
\[\begin{array}{ccccccc}
    d_1 	&=&			c_1 X_{1,1} &+& \dotsb 	&+& c_m X_{1,m} \\
    \vdots 	&& 			\vdots		&&	\ddots  &&	\vdots \\
    d_n 	&=& 		c_1 X_{n,1} &+& \dotsb 	&+& c_m X_{n,m}
\end{array}\]
with variables $X_{i,j}$ for $i = 1,\dots,n$ and $j = 1,\dots,m$ and the additional constraint
\[\bigwedge_{j = 1}^m \left(x_j \in \varset_+ \rightarrow \sum_{i = 1}^n X_{i,j} \geq 1\right).\]
\end{definition}
Each solution $\alpha : X \rightarrow \mathbb{N}$ to these equations corresponds to a substitution
$\sigma_\alpha$ where $\sigma_\alpha(x_j) = M_{_\alpha,j}$ for the smallest multiset $M_{_\alpha,j}$
with $\multiplicity_{M_{_\alpha,j}}(s_i) = \alpha(X_{i,j})$ for $i = 1,\dots,n$ and $j = 1,\dots,m$.

Extensive research has been done on solving linear diophantine equations and linear diophantine
equation systems \citep{Weinstock1960,Bond1967,Lambert1988,Clausen1989,Aardal2000}. Our equation
system is however not a real system, because the equations are disjoint and can be solved
independently. Also, since we are only interested in the non-negative solutions, these solutions can
be found more easily. We implemented an algorithm that is an adaptation of the
algorithm used in SymPy \citep[in sympy.solvers.diophantine]{SymPy}. The algorithm recursively
reduces any linear diophantine equation to equations of the form $ax + by = d$ that can be solved
efficiently with the Extended Euclidian algorithm \citep[see 2.107]{Menezes1996}. Then the solutions
for those can be combined into a solution for the original equation.

In all practical applications that we have considered, both the multiplicities of the variables
and the multiplicities of terms in commutative patterns are small. Hence, for a sequence variable
equation $c_1 x_1 + \dots c_n x_n = c$, all coefficients $c_i$ and $c$ are usually small.
Furthermore, the number $n$ of distinct sequence variables in a pattern is also typically
small. Hence, the number of solutions for these equations is usually also small and we can
cache the results instead of recomputing them every time. The number of distinct cases can be
reduced by sorting the coefficients and hence the storage for the caching can be reduced. This makes
the time cost of enumerating all distributions of the terms onto the sequence variables effectively
linear per distribution.

For each sequence variable substitution that is enumerated by the above method, the compatibility
with the existing substitutions is checked and the resulting union is yielded as are result if they
are compatible.
\chapter{Many-to-One Pattern Matching} \label{chp:many}

For many problems there is a fixed set of patterns which are all matched against different subjects repeatedly. While this can be solved by matching each pattern separately there is room for optimization. In this chapter we present data structures and algorithms to do many-to-one matching more efficiently.

\begin{definition}[Many-to-one Pattern Matching]
Many-to-one pattern matching is the process of matching a single subject against multiple patterns simultaneously.
\end{definition}

\section{Discrimination Net}

For syntactic patterns, \glspl{DN}, also known as left-to-right tree automata, have proven to be fast for many-to-one matching
\citep{Christian1993,Graef1991,Nedjah1997}. In this section we discuss existing kinds of \glspl{DN}
and propose extensions which can match a wider set of patterns.

\subsection{Syntactic Discrimination Net}
\glspl{DN} are a specialized form of finite automata which can be either deterministic or non-deterministic.
A deterministic \gls{DN} can match a subject of size $n$ against all its patterns in $\mathcal{O}(n)$.
However, its size can potentially grow exponentially with the number of patterns.
Non-deterministic \glspl{DN} take up smaller space ($\mathcal{O}(m)$ where $m$ is the sum of pattern lengths),
but also have a worst case runtime of $\mathcal{O}(m)$. We mostly focus on deterministic \glspl{DN} for syntactic matching,
because for the cases that we considered they offer the best performance.

\cite{Christian1993} also proposed the use of a \emph{flatterm} representation instead of a tree based one. This representation is based on the preorder
traversal of the tree and hence speeds up algorithms relying mostly on that. The flatterm representation of a term is a word formed by the symbols as they
are encountered in preorder traversal of the tree structure, \eg the flatterm for $f(a, g(b, c))$ would be $f\,a\,g\,b\,c$. This notation as
well as \glspl{DN} in the literature only support function symbols with fixed arity. In the implementation, additional pointers are
used to preserve some of the tree structure while providing fast preorder traversal.

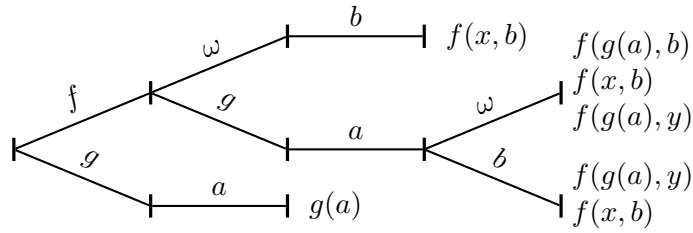
\begin{figure}[h]
    \centering
    \begin{tikzpicture}[
        grow=right,
        sloped,
        thick,
        level 2/.style={sibling distance=15mm},
        level distance=18mm
    ]
        \node[st2b] {}
            child {
                node[st2b] {}
                child {
                    node[st2b] {}
                    node [label=right:$g(a)$] {}
                    edge from parent
                    node[above] {$a$}
                }
                edge from parent
                node[above] {$g$}
            }
            child {
                node[st2b] {}
                child {
                    node[st2b] {}
                    child {
                        node[st2b] {}
                        child {
                            node[st2b] {}
                            node [label={[right,align=left]$f(g(a), y)$ \\ $f(x, b)$}] {}
                            edge from parent
                            node[above] {$b$}
                        }
                        child {
                            node[st2b] {}
                            node [label={[right,align=left]$f(g(a), b)$ \\ $f(x, b)$ \\ $f(g(a), y)$}] {}
                            edge from parent
                            node[above] {$\omega$}
                        }
                        edge from parent
                        node[above] {$a$}
                    }
                    edge from parent
                    node[above] {$g$}
                }
                child {
                    node[st2b] {}
                    child {
                        node[st2b] {}
                        node [label=right:{$f(x, b)$}] {}
                        edge from parent
                        node[above] {$b$}
                    }
                    edge from parent
                    node[above] {$\omega$}
                }
                edge from parent
                node[above] {$f$}
            };
    \end{tikzpicture}
    \caption{Example of a Discrimination Net}
    \label{fig:disc_net}
\end{figure}

Usually, the patterns are treated as if they were linear, \ie variable symbols are all treated the same and replaced with $\omega$ which acts as a general placeholder.
Therefore, in case of non-linear patterns, the equivalence of variable replacements has to be checked in an extra step.
\autoref{fig:disc_net} gives an example of a \gls{DN} for the pattern set $\{f(g(a), b),\allowbreak f(x, b), f(g(a), y), g(a)\}$.
The leafs of the \gls{DN} are labeled with the matched patterns. Matching is done by interpreting
the tree as an automaton and run it on the flatterm representation of the subject. The initial state
is the root of the tree and the leafs are the final states. An $\omega$ transition covers
all symbols for which there is no other outgoing transition yet. Note that an $\omega$ transition skips the whole
subterm if a function symbol is read. If the automaton accepts, its final state determines the set of matching patterns. Otherwise, none of the patterns match.

\begin{figure}[h]
    \centering
    \begin{tikzpicture}[
        grow=right,
        sloped,
        thick,
        level 2/.style={level distance=5mm},
        level 3/.style={level distance=10mm},
        level 4/.style={level distance=5mm},
        level 5/.style={level distance=15mm},
        level distance=15mm
    ]
        \node[st2b,label=below:$q_0$] {}
            child {
                node[st2b,label=below:$q_1$] {}
                child {
                    node[minimum height=0pt, minimum width=0pt, inner sep=0pt] {}
                    child {
                        node[minimum height=0pt, minimum width=0pt, inner sep=0pt] {}
                        child {
                            node[st2b,solid,label=below:$q_{n-1}$] {}
                            child {
                                node[st2b,label=below:$q_n$] {}
                                node [label={[right,align=left]$p$}] {}
                                edge from parent
                                node[above] {$s_n$}
                            }
                            edge from parent[solid]
                        }
                        edge from parent[dotted]
                    }
                    edge from parent
                }
                edge from parent
                node[above] {$s_1$}
            };
    \end{tikzpicture}
    \caption{Discrimination Net for a Single Pattern}
    \label{fig:single_disc_net}
\end{figure}
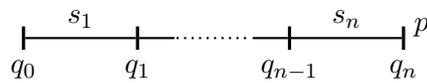

The construction of a deterministic \gls{DN} is straight-forward and can be done by using product automaton construction.
For every pattern $p = s_1 \dots s_n$ in flatterm form we construct a simple linear net like in \autoref{fig:single_disc_net}.
These can be combined with standard DFA product construction while merging the pattern sets of each state.
Other (incremental) construction methods are described in \cite{Christian1993,Graef1991,Nedjah1997}.
While theoretically the resulting \gls{DN} can have exponentially many
nodes\footnote{Consider the pattern set $P_n := \{\, f(x_1, \dots, x_{i-1}, a, x_{i+1}, \dots, x_n) \mid 1 \leq i \leq n\ \,\}$ which contains $n$ patterns
and results in $2^n$ distinct leaf nodes as all patterns can match independently of each other.}, for many practical pattern sets this does not
pose a problem (for an example, see \autoref{fig:vsdn:size}).

However, we cannot use this \gls{DN} to match variadic functions as it cannot distinguish between different terms having the same flatterm representation.
Consider the previous example but now $f$ and $g$ are variadic function symbols. Because \eg $f(g(a), b)$ and $f(g(a, b))$ have the same flatterm representation
$f\,g\,a\,b$, they are treated the same by the \gls{DN}.
In the following section, a generalized version of the \gls{DN} which supports variadic function symbols.

\subsection{Variadic Syntactic Discrimination Net}

In order to use variadic function symbols with \glspl{DN}, we include a special symbol ``$\fend$'' denoting the end of a compound term in the preorder traversal.
We add this symbol as the last argument to every compound term, \eg $f(a, g(b), c)$ is transformed to $f(a, g(b, \fend), c, \fend)$ resp.
$f\,a\,g\,b\,\fend\,c\,\fend$ as a flatterm instead.

\begin{definition}[Variadic Syntactic Discrimination Net] \label{def:VSDN}
    A \gls{VSDN} $\mathfrak{D}$ is defined similar to a DFA as $\mathfrak{D} = (Q, \Sigma, \delta, q_0, F)$ where
    \begin{itemize}
        \item $Q$ is a finite set of states,
        \item $\Sigma = \funcset \cup \{ \fend \}$ is the finite alphabet,
        \item $\delta: Q \times \Sigma \rightarrow Q \times \{ 0, 1 \}$ is the transition function where the second result component is $1$ iff an $\omega$ transition was used,
        \item $q_0 \in Q$ is the initial state, and
        \item $F: Q \rightarrow 2^{\termset(\funcset, \varset)}$ which maps each (leaf) state to a set of patterns it covers.
    \end{itemize}
    A \emph{run} of $\mathfrak{D}$ on a term $t$ is a sequence of states with positions $\rho = (q_1, \nu_1) \dots (q_n, \nu_n)$ such that for every $i = 2 ,\dots,n$ we have either
    \begin{itemize}
        \item $\delta(q_{i-1}, h) = (q_i, 0)$ where $h = head(t|_{\nu_{i-1}})$ and $\nu_i = next(t, \nu_{i-1})$, or
        \item $\delta(q_{i-1}, h) = (q_i, 1)$ and $\nu_i = skip(t, \nu_{i-1})$.
    \end{itemize}
    A pattern $p$ \emph{matches} a subject $s$ in $\mathfrak{D}$ iff there exists a run $\rho = (q_0, \epsilon) \dots (q_{n-1}, \nu_{n-1})\allowbreak (q_n, \top)$ on $s$ where $q_0$ is
    the initial state, $\nu_i \in \pos(p)$ for $i = 0 ,\dots,n-1$, and $p \in F(q_n)$.
\end{definition}

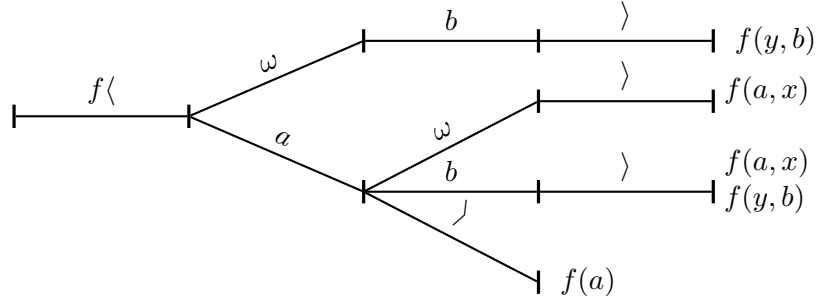
\begin{figure}[h]
    \centering
    \begin{tikzpicture}[
        grow=right,
        sloped,
        thick,
        level 2/.style={sibling distance=20mm},
        level 3/.style={sibling distance=12mm},
        level distance=23mm
    ]
        \node[st2b] {}
            child {
                node[st2b] {}
                child {
                    node[st2b] {}
                    child {
                        node[st2b] {}
                        node [label=right:$f(a)$] {}
                        edge from parent
                        node[above] {$\fend$}
                    }
                    child {
                        node[st2b] {}
                        child {
                            node[st2b] {}
                            node [label={[right,align=left]$f(a, x)$ \\ $f(y, b)$}] {}
                            edge from parent
                            node[above] {$\fend$}
                        }
                        edge from parent
                        node[above] {$b$}
                    }
                    child {
                        node[st2b] {}
                        child {
                            node[st2b] {}
                            node [label={[right,align=left]$f(a, x)$}] {}
                            edge from parent
                            node[above] {$\fend$}
                        }
                        edge from parent
                        node[above] {$\omega$}
                    }
                    edge from parent
                    node[above] {$a$}
                }
                child {
                    node[st2b] {}
                    child {
                        node[st2b] {}
                        child {
                            node[st2b] {}
                            node [label=right:{$f(y, b)$}] {}
                            edge from parent
                            node[above] {$\fend$}
                        }
                        edge from parent
                        node[above] {$b$}
                    }
                    edge from parent
                    node[above] {$\omega$}
                }
                edge from parent
                node[above] {$f\fstart$}
            };
    \end{tikzpicture}
    \caption{Example of a Variadic Syntactic Discrimination Net}
    \label{fig:VSDN}
\end{figure}

In \autoref{fig:VSDN}, a \gls{VSDN} net for the pattern set $\{f(a, x), f(a),\allowbreak f(y, b)\}$ is shown.
Note that $f$ is a variadic function symbol here and we added a ``$\fstart$'' to function symbols to distinguish them from constant symbols.

The \gls{VSDN} could be extended to support sequence variables as well by adding $\omega$-loops to states.
By interpreting the \gls{VSDN} as a NFA, one can use $\epsilon$-transitions and use powerset construction to get a minimal \gls{VSDN}
for a pattern with sequence variables. These individual \glspl{VSDN} can be combined using product construction.

Nonetheless, experiment with sequence variables have shown to increase the size of the \gls{VSDN} beyond the
point where it is beneficial (see \autoref{fig:vsdn:size_ctx}). A non-deterministic approach is more suitable to deal with sequence variables and associativity.
In the next section, such an approach is discussed.

\subsection{Associative Discrimination Net}

In this section we propose a new type of non-deterministic discrimination net that supports sequence variables and associative function symbols. We can later use
multiple layers of this \gls{ADN} to perform commutative matching as well. Furthermore, in contrast to most existing work, non-linear patterns are handled by
the \gls{ADN} itself, because that enables earlier backtracking when a variable mismatch occurs.

\begin{definition}[Associative Discrimination Net]
    \label{def:ADN}
    An \gls{ADN} $\mathfrak{A}$ is a non-deterministic extension of a \gls{VSDN} which is defined as $\mathfrak{A} = (Q, \Sigma, \delta, q_0, F)$ where
    \begin{itemize}
        \item $Q$ is a finite set of states,
        \item $\Sigma = \funcset \cup \varset \cup \{ \fend \}$ is the finite alphabet,
        \item $\delta \subseteq Q \times \Sigma \times Q$,
        \item $q_0 \in Q$ is the initial state, and
        \item $F: Q \rightarrow \termset(\funcset, \varset) \cup \{\emptyset\}$ which maps each (leaf) state to the patterns it accepts.
    \end{itemize}
    A \emph{run} of $\mathfrak{A}$ on a term $t$ is a sequence of states with positions and substitutions
    $\rho = (q_1, \nu_1, \sigma_1) \dots (q_n, \nu_n, \sigma_n)$ such that for every $i = 2, \dots, n$ we have that $\sigma_{i-1} \fcompatible \sigma'_i$,
    $\sigma_i = \sigma_{i-1} \funion \sigma'_i$, and either
    \begin{itemize}
        \item $(q_{i-1}, h, q_i) \in \delta$ where $h = head(t|_{\nu_{i-1}})$,
              $\nu_i = next(t, \nu_{i-1})$ and
              $\sigma'_i := \emptyset$, or
        \item $(q_{i-1}, x, q_i) \in \delta$ for some $x \in \varset$,
              $\nu_i = skip(t, \nu_{i-1})$ and
              $\sigma'_i = \{ x \mapsto t|_{\nu_{i-1}} \}$, or
        \item $(q_{i-1}, x, q_i) \in \delta$ for some $x \in \varset_*$,
              $\nu_i = \nu_{i-1}$ and
              $\sigma'_i = \{ x \mapsto \epsilon \}$, or
        \item $(q_{i-1}, x, q_i) \in \delta$ for some $x \in \varset_+ \cup \varset_*$,
              $\nu_{i-1} = \nu_p k < \nu_p j \in \pos(t)$ for some $\nu_p \in \pos(t)$ and $k, j \in \mathbb{N}$,
              $skip(t, \nu_p j) = \nu_i$ and
              $\sigma'_i = \{ x \mapsto (t|_{\nu_p k}, \dots, t|_{\nu_p j}) \}$, or
        \item $(q_{i-1}, x, q_i) \in \delta$ for some $x \in \varset_0$,
              $\nu_{i-1} = \nu_p k < \nu_p j \in \pos(t)$ for some $\nu_p \in \pos(t)$ and $k, j \in \mathbb{N}$, $f = head(t|_{\nu_p})$ is associative,
              $skip(t, \nu_p j) = \nu_i$ and
              $\sigma'_i = \{ x \mapsto f(t|_{\nu_p k}, \dots, t|_{\nu_p j}) \}$.
    \end{itemize}
    A pattern $p$ \emph{matches} a subject $s$ with a substitution $\sigma$ in $\mathfrak{A}$ iff there exists a
    run $\rho = (q_0, \epsilon, \emptyset) \dots (q_{n-1}, \nu_{n-1}, \sigma_n) (q_n, \top, \sigma)$ on $s$ where $q_0$ is
    the initial state, $\nu_i \in \pos(p)$ for $i = 0 ,\dots,n-1$, and $p = F(q_n)$.
\end{definition}

Furthermore, to support guards directly attached to subterms, we can extend the definition of a run:
In addition to allowing a transition $(q_{i-1}, s, q_i) \in \delta$ for some $s$, we we also allow
$(q_{i-1}, s\ \mathbf{if}\ \varphi, q_i) \in \delta$ for some $s$ and $\varphi$. For those guarded
transitions it is required that $\varphi(\sigma_i) = true$ for them to be included in a valid run.

\begin{figure}[h]
    \centering
    \begin{tikzpicture}[
        grow=right,
        sloped,
        thick,
        level 2/.style={sibling distance=20mm},
        level 3/.style={sibling distance=12mm},
        level distance=23mm
    ]
        \node[st2b,label=above:0] {}
            child {
                node[st2b,label=above:1] {}
                child {
                    node[st2b,label=above:2] {}
                    child {
                        node[st2b,label=above:3] {}
                        node [label={[right,align=left]$f(a)$}] {}
                        edge from parent
                        node[above] {$\fend$}
                    }
                    child {
                        node[st2b,label=above:4] {}
                        child {
                            node[st2b,label=above:5] {}
                            node [label={[right,align=left]$f(a, x^*)$}] {}
                            edge from parent
                            node[above] {$\fend$}
                        }
                        edge from parent
                        node[above] {$x^*$}
                    }
                    edge from parent
                    node[above] {$a$}
                }
                child {
                    node[st2b,label=above:6] {}
                    child {
                        node[st2b,label=above:7] {}
                        child {
                            node[st2b,label=above:8] {}
                            node [label=right:{$f(y, b)$}] {}
                            edge from parent
                            node[above] {$\fend$}
                        }
                        edge from parent
                        node[above] {$b$}
                    }
                    edge from parent
                    node[above] {$y$}
                }
                edge from parent
                node[above] {$f\fstart$}
            };
    \end{tikzpicture}
    \caption{Example of an Associative Discrimination Net}
    \label{fig:ADN}
\end{figure}
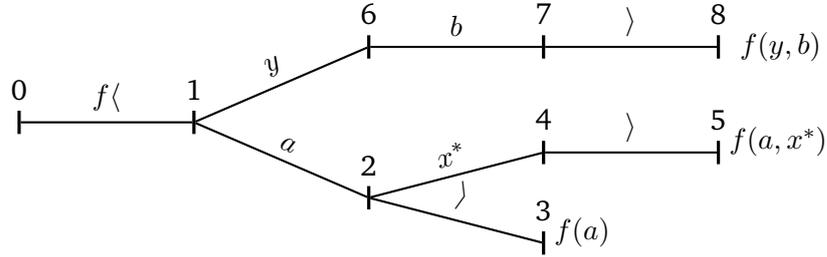

\autoref{fig:ADN} shows the same example as \autoref{fig:VSDN}, but as non-deterministic version with non-linear variables.
Given the subject $f(a, b)$, both $f(a, x^*)$ and $f(y, b)$ match. For $f(a, x^*)$ we get the run
\[(0, \epsilon, \emptyset) (1, 1, \emptyset) (2, 2, \emptyset) (4, 3, \sigma_1) (5, \top, \sigma_1)\]
with $\sigma_1 = \{ x^* \mapsto (b) \}$. For $f(y, b)$ we have the run
\[(0, \epsilon, \emptyset) (1, 1, \emptyset) (6, 2, \sigma_2) (7, 3, \sigma_2) (8, \top, \sigma_2)\]
with $\sigma_2 = \{ y \mapsto a \}$.
When only syntactic pattern are used, the \gls{ADN} specializes to a non-deterministic \gls{VSDN}. The advantage of the \gls{ADN} emerges when patterns
share some common structure and contain sequence variables, as the non-deterministic partitioning of terms among sequence variables needs only be done once.
Because the partitioning is the most expensive part of the (non-commutative) matching, this improves the efficiency of the matching.
In the worst case matching with an \gls{ADN} is equivalent to one-to-one matching, because the
decision branches in Algorithm \ref{alg:match_one_to_one} and Algorithm \ref{alg:match_function}
map directly to the structure of the \gls{ADN}.

\begin{algorithm}[h]
\caption{Adding a pattern to an \gls{ADN}}
\label{alg:addpattern}
\begin{algorithmic}[1]
    \Require{\par%
        Discrimination net $\mathfrak{A} = (Q, \Sigma, \delta, q_0, F)$ and
        pattern in flatterm representation $p = s_1 \dots s_n$ with $s_i \in \Sigma$ for $i = 1, \dots, n$.
    }
    \algrule
    \Function{addPattern}{$\mathfrak{A}, p$}
        \Let{$q$}{$q_0$}
        \ForAll{$i \in 1,\dots,n$}
            \If{$\exists q' \in Q: (q, s_i, q') \in \delta$}
                \Let{$q$}{$q'$}
            \Else
                \State Create new state $q'$
                \Let{$Q$}{$Q \cup \{q'\}$}
                \Let{$\delta$}{$\delta \cup \{(q, s_i, q')\}$}
                \Let{$q$}{$q'$}
            \EndIf
        \EndFor
        \Let{$F$}{$F[q \mapsto F(q) \cup \{p\}]$}
    \EndFunction
\end{algorithmic}
\end{algorithm}

\begin{algorithm}[h]
\caption{Matching with an \gls{ADN}}
\label{alg:match}
\begin{algorithmic}[1]
    \Require{\par%
        Discrimination net $\mathfrak{A} = (Q, \Sigma, \delta, q_0, F)$,
        subject $s \in \termset$, a state $q \in Q$, a position $\nu \in \pos(s)$
        and a substitution $\sigma \in \substs$.
    }
    \algrule
    \Function{ADNMatch}{$\mathfrak{A}, s, q, \nu, \sigma$}
        \If{$\nu = \top$}
            \Return{$\{(F(q), \sigma)\}$}
        \Else
            \Let{$m$}{$\emptyset$}
            \Let{$h$}{$head(s|_\nu)$}
            \IndentedLineComment{Symbol transition}{2}
            \If{$\exists q' \in Q, \exists \varphi: (q, h\ \mathbf{if}\ \varphi, q') \in \delta$ and $\varphi(\sigma) = true$}
                \Let{$m$}{$m \cup \textsc{ADNMatch}(\mathfrak{A}, s, q', next(s, \nu), \sigma)$}
            \EndIf
            \ForAll{$(q, x\ \mathbf{if}\ \varphi, q') \in \delta$ for some $x \in \varset$, $q' \in Q$ and $\varphi$}
                \IndentedLineComment{Variable transition matching a single term}{3}
                \If{$\sigma \fcompatible \{ x \mapsto s|_\nu \}$ and $\varphi(\sigma) = true$}
                    \Let{$m$}{$m \cup \textsc{ADNMatch}(\mathfrak{A}, s, q', skip(s, \nu), \sigma \funion \{ x \mapsto s|_\nu \})$}
                \EndIf
                \IndentedLineComment{Star variable transition matching empty sequence}{3}
                \If{$x \in \varset_*$ and $\sigma \fcompatible \{ x \mapsto \epsilon \}$ and $\varphi(\sigma) = true$}
                    \Let{$m$}{$m \cup \textsc{ADNMatch}(\mathfrak{A}, s, q', \nu, \sigma \funion \{ x \mapsto \epsilon \})$}
                \EndIf
                \If{$\nu = \nu_p k$ for some $\nu_p \in Pos(s)$ and $k \in \mathbb{N}$}
                    \IndentedLineComment{Parent term has position $\nu_p$ and current term is argument no. $k$}{4}
                    \Let{$f$}{$head(s|_{\nu_p})$}
                    \If{$x \in \varset_+ \cup \varset_*$ or $f$ is associative}
                        \Let{$j$}{$\argmax_{j \in \mathbb{N}} \nu_p j \in Pos(s)$} \Comment{$j$ is no. of last argument}
                        \ForAll{$i = k+1 \dots j$}
                            \Let{$t$}{$(s|_{\nu_p k}, \dots, s|_{\nu_p i})$}
                            \If{$x \in \varset_0$ and $f$ is associative}
                                \Let{$t$}{$f(t)$}  \Comment{Wrap regular variable in assoc. function}
                            \EndIf
                            \IndentedLineComment{Sequence variable transition matching}{6}
                            \If{$\sigma \fcompatible \{ x \mapsto t \}$ and $\varphi(\sigma) = true$}
                                \Let{$m$}{$m \cup \textsc{ADNMatch}(\mathfrak{A}, s, q', skip(s, \nu_p i), \sigma \funion \{ x \mapsto t \})$}
                            \EndIf
                        \EndFor
                    \EndIf
                \EndIf
            \EndFor
            \State\Return{$m$}
        \EndIf
    \EndFunction
\end{algorithmic}
\end{algorithm}

The construction of the \gls{ADN} can be done by iteratively adding new patterns to it starting from the initial
state as outlined in Algorithm~\ref{alg:addpattern}.
In order to maximize the potential similarity between patterns, their variables
can be renamed to a position\hyphen{}based canonical schema. We replace every variable
$x$ with a name based on its first occurrence in $t$, \ie replace $x$ with a new variable
$x_{p_1(x)}$ for every $x \in \vars(t)$ where
\[p_1(x) := \min \{p \in \pos(t) \mid t|_p = x\}.\]
As an example, $f(x, y, g(z))$ would become $f(x_{1}, x_{2}, g(x_{3\,1}))$. This way, terms with
similar structure but differently named variables can still be combined in the discrimination
net. Note that this does not change that regular and sequence variables still need to be handled
differently. While $f(x)$ and $f(y)$ are treated as equivalent by the \gls{ADN},
$f(x^+)$ has different transitions and a distinct final state from the other two.
After a match is found, the renaming of the variables is reverted to obtain the correct substitution.

The matching with an \gls{ADN} is done be trying all possible runs with backtracking.
For simplicity, the pseudo code for the algorithm assumes that every subterm has a guard.
If a term does not have a guard, we replace it with the equivalent term that has a $true$ guard,
\eg $f(x, a)$ would become $f(x\ \mathbf{if}\ true, a\ \mathbf{if}\ true)\ \mathbf{if}\ true$.
This reduces the number of cases that need to be distinguished for the pseudo code, but in an
implementation guards can easily be separated.
In Algorithm \ref{alg:match}, the pseudo code for a many-to-one match function is listed.
It yields the set of all matching pairs of patterns and substitutions, \ie
$\textsc{ADNMatch}(\mathfrak{A}, s, q_0, \epsilon, \emptyset)$ yields all matches for a subject $s$.

\section{Many-to-one Matcher}

In order to use the \gls{ADN} for patterns with commutative function symbols as well, we handle them in a separate step.
For that we add a special state for every commutative subpattern of a pattern and use another \gls{ADN} to match its children.
Finally, we find matches for the whole commutative subpattern by finding maximum matchings in a bipartite
graph constructed from the matches for the subject's subterms in the \gls{ADN}. This is similar to the approach of the AC discrimination nets described in \cite{Bachmair1995},
but we perform the full non-linear matching directly at a commutative symbol state instead of just using it as a filter.

\pagebreak
\subsection{Multilayer Discrimination Net}

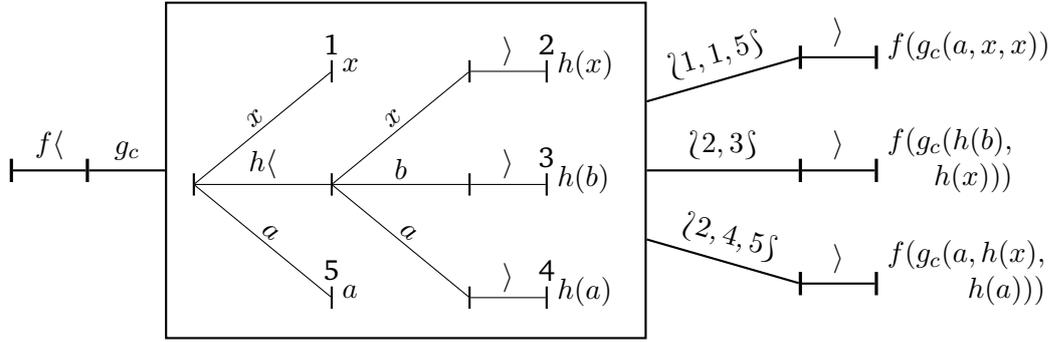
\begin{figure}[h]
    \centering
    \begin{tikzpicture}[
        grow=right,
        sloped,
        thick,
        level 1/.style={level distance=10mm}
    ]
        \node[st2b] {}
            child {
                node[st2b] (main) {}
                child {
                    node[draw=black,right=1cm of main,inner sep=10pt] (commutative) {
                        \begin{tikzpicture}[every node/.style={inner sep=2pt},level 3/.style={level distance=18mm},level 5/.style={level distance=10mm}]
        \node[st2b,inner sep=0pt] {}
            child {
                node[st2b,label=5] {}
                node [label={[right,align=left]$a$}] {}
                edge from parent
                node[above] {$a$}
            }
            child {
                node[st2b] {}
                child {
                    node[st2b] {}
                    child {
                        node[st2b,label=4] {}
                        node [label={[right,align=left]$h(a)$}] {}
                        edge from parent
                        node[above] {$\fend$}
                    }
                    edge from parent
                    node[above] {$a$}
                }
                child {
                    node[st2b] {}
                    child {
                        node[st2b,label=3] {}
                        node [label={[right,align=left]$h(b)$}] {}
                        edge from parent
                        node[above] {$\fend$}
                    }
                    edge from parent
                    node[above] {$b$}
                }
                child {
                    node[st2b] {}
                    child {
                        node[st2b,label=2] {}
                        node [label={[right,align=left]$h(x)$}] {}
                        edge from parent
                        node[above] {$\fend$}
                    }
                    edge from parent
                    node[above] {$x$}
                }
                edge from parent
                node[above] {$h\fstart$}
            }
            child {
                node[st2b,label=1] {}
                node [label={[right,align=left]$x$}] {}
                edge from parent
                node[above] {$x$}
            };
                        \end{tikzpicture}
                    }
                    child {
                        node[st2b,right=2cm of commutative,shift={(0, 1.5cm)}] {}
                        child {
                            node[st2b]  {}
                            node [label={[right,align=left]$f(g_c(a, x, x))$}] {}
                            edge from parent
                            node[above] {$\fend$}
                        }
                        edge from parent
                        node[above] {$\multiset{1, 1, 5}$}
                    }
                    child {
                        node[st2b,right=2cm of commutative,shift={(0, -1.5cm)}] {}
                        child {
                            node[st2b]  {}
                            node [label={[right,align=right]$f(g_c(a, h(x),$\\$h(a)))$}] {}
                            edge from parent
                            node[above] {$\fend$}
                        }
                        edge from parent
                        node[above] {$\multiset{2, 4, 5}$}
                    }
                    child {
                        node[st2b,right=2cm of commutative]  {}
                        child {
                            node[st2b]  {}
                            node [label={[right,align=right]$f(g_c(h(b),$\\$h(x)))$}] {}
                            edge from parent
                            node[above] {$\fend$}
                        }
                        edge from parent
                        node[above] {$\multiset{2, 3}$}
                    }
                    edge from parent
                    node[above] {$g_c$}
                }
                edge from parent
                node[above] {$f\fstart$}
            };
        \end{tikzpicture}
        \caption{Example of a Multilayer Discrimination Net}
        \label{fig:MLDN}
\end{figure}

In \autoref{fig:MLDN}, a \gls{MLDN} for the pattern set \[\{f(g_c(a, x, x)),\allowbreak f(g_c(a, h(x), h(a))), f(g_c(h(b), h(x)))\}\] is shown.
For the $g_c$ compound term, a special state containing another \gls{ADN} is added. The outgoing transitions from that state are labeled with
a \emph{requirement multiset} that defines which subpatterns are required to match in order for the whole $g_c$ term to match. These sets are later used to
construct bipartite graphs to find all valid matches.

As an example, consider the subject $f(g_c(a, h(a), h(a)))$. $a$ and $h(a)$ are both run through
the inner \gls{ADN}. $a$ matches both $a$ and $x$ (with a \emph{match set} $\{1, 5\}$). $h(a)$ matches $x$, $h(x)$,
and $h(a)$ (with a match set $\{1, 2, 4\}$).

To exclude patterns which cannot match from further
processing, we check multiset union of the match sets against each requirement multiset.
The multiplicity of each subpattern is used when building the multiset union. In the above example
the multiset union would be
\[P := \{1, 5\} \uplus \{1, 2, 4\} \uplus \{1, 2, 4\} = \multiset{ 1, 1, 1, 2, 2, 4, 4, 5 }.\]
Note that the match set for $h(a)$ is included two times since the term appears two times in the
subject.
Hence, $f(g_c(h(b), h(x)))$ cannot match the subject, because $\multiset{2, 3} \not\subseteq P$.
For the other two patterns, we have both $\multiset{1, 1, 5} \subseteq P$ and $\multiset{2, 4, 5} \subseteq P$
so both could match and need to be processed further.

To show that this check is not sufficient for determining whether a pattern matches
consider the subject $f(g_c(a, a, h(a)))$ with its multiset union $P' = \multiset{ 1, 1, 1, 2,\allowbreak 4, 5, 5 }$.
Despite $\multiset{1, 1, 5} \subseteq P$, there is no match for $f(g_c(a, x, x))$ because
$\{ x \mapsto a \}$ and $\{ x \mapsto h(a) \}$ are not compatible. Similarly, even though
$\multiset{2, 4, 5} \subseteq P$, there is no match for $f(g_c(a, h(x), h(a)))$ because multiple
both $h(x)$ and $h(a)$ would have to be covered by the single term $h(a)$.

Therefore, confirming if one of the patterns actually matches requires a more involved check. This
is solved by constructing bipartite graphs from the match sets and finding maximum bipartite matchings.
However, the filtering step described above can reduce the number of more expensive bipartite
matching problems that need to be solved.

\subsection{Bipartite Graphs}

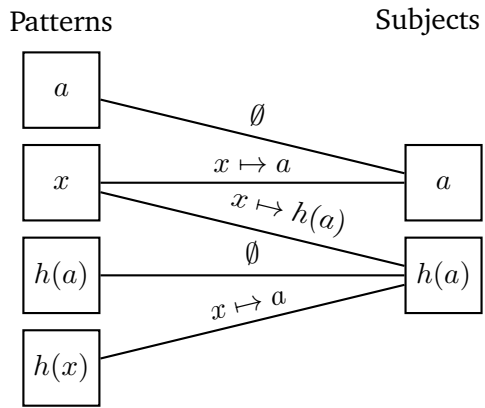
\begin{figure}[h]
    \centering
    \begin{tikzpicture}[sloped,thick,term/.style={draw,rectangle,minimum size=1cm}]
         \node[term] (p_a) {$a$};
         \node[term,below=0.2cm of p_a] (p_x) {$x$};
         \node[term,below=0.2cm of p_x] (p_ha) {$h(a)$};
         \node[term,below=0.2cm of p_ha] (p_hx) {$h(x)$};
         \node[term,right=4cm of p_x] (s_a) {$a$};
         \node[term,below=0.2cm of s_a] (s_ha) {$h(a)$};
         \node[above=0.4cm of p_a,anchor=center] (patterns) {Patterns};
         \node[right=4cm of patterns,anchor=center] {Subjects};

         \draw (p_x) -- (s_a) node[midway,above] {$x \mapsto a$};
         \draw (p_x) -- (s_ha) node[midway,above,pos=0.6] {$x \mapsto h(a)$};
         \draw (p_a) -- (s_a) node[midway,above] {$\emptyset$};
         \draw (p_ha) -- (s_ha) node[midway,above] {$\emptyset$};
         \draw (p_hx) -- (s_ha) node[midway,above] {$x \mapsto a$};
    \end{tikzpicture}
    \caption{Full Bipartite Graph}
    \label{fig:bipartite}
\end{figure}

We construct a bipartite graph where one set of nodes contains the distinct subpatterns
and the other one consists of the distinct terms of the subject. Then the matches found by running
the terms through the \gls{ADN} induce the edges, \ie a subpattern node and a term node are
connected by an edge iff the subpattern matches the term. We label the edge with the set of match
substitutions. The resulting bipartite graph for the previous example with subject
$f(g_c(a, h(a), h(a)))$ is shown in \autoref{fig:bipartite}.

For matching the whole commutative pattern we use the subgraph induced by only including the nodes
of this particular pattern's subpatterns. Furthermore, for terms in the patterns or subject which
occur multiple times, we need to duplicate the corresponding nodes, \eg for the above subject
example the $h(a)$ node would need to be duplicated in the graph that is used to find the maximum
matchings.

\begin{figure}[h]
    \centering
    \begin{tikzpicture}[sloped,thick,term/.style={draw,rectangle,minimum size=1cm}]
         \node[term,label=left:1] (p_x1) {$x$};
         \node[term,below=0.5cm of p_x1,label=left:2] (p_x2) {$x$};
         \node[term,below=0.5cm of p_x2,label=left:3] (p_a) {$a$};
         \node[term,right=4cm of p_x1,label=right:1] (s_ha1) {$h(a)$};
         \node[term,below=0.5cm of s_ha1,label=right:2] (s_ha2) {$h(a)$};
         \node[term,below=0.5cm of s_ha2,label=right:3] (s_a) {$a$};
         \node[above=0.4cm of p_x1,anchor=center] (patterns) {Patterns};
         \node[right=4cm of patterns,anchor=center] {Subjects};

         \draw (p_x1) -- (s_ha1) node[midway,above] {1};
         \draw (p_x1) -- (s_ha2) node[above,pos=0.3] {2};
         \draw (p_x2) -- (s_ha1) node[above,pos=0.65] {3};
         \draw (p_x2) -- (s_ha2) node[above,pos=0.7] {4};
         \draw (p_x1) -- (s_a) node[above,pos=0.8] {5};
         \draw (p_x2) -- (s_a) node[above,pos=0.45] {6};
         \draw (p_a) -- (s_a) node[above,pos=0.3] {7};
    \end{tikzpicture}
    \caption{Induced Bipartite Graph}
    \label{fig:bipartite2}
\end{figure}
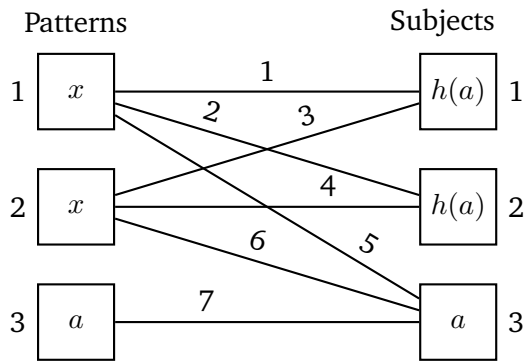

In \autoref{fig:bipartite2} the induced graph for the pattern $f(g_c(a, x, x))$ and the subject
$f(g_c(a,\allowbreak h(a), h(a)))$ is shown. The substitutions on the edges have been left out here for
readability, but they are the same as in the original graph. Instead we have labeled every edge
with a number for later reference. All nodes are also labeled with consecutive natural numbers to
distinguish duplicated nodes. Note that patterns and subject are counted
separately. The problem of finding all matches for a commutative pattern and subject can be reduced
to finding all maximum matchings in the induced bipartite graph and checking that the substitutions
of the matching edges are compatible. The union of the matching edge substitutions then yields a valid
match for the whole commutative term.

However, because we duplicated some nodes, we have some redundant matchings that result in the same
substitution. Here we can use the additional number labels on the nodes to define a canonical
maximum matching for every distinct match and disregard non-canonical maximum matchings:
\begin{definition}[Canonical Matching]
Let $P$ be a set of patterns and $S$ be a set of subjects. A matching $M \subseteq E$ on an induced
bipartite graph $B$ with (directed) edge set $E \subseteq (P \times \mathbb{N}) \times (S \times \mathbb{N})$
over $P$ and $S$ is canonical iff $\forall (p, n_p, s, n_s) \in M, \forall (p', n_p', s', n_s') \in M:
(s = s' \wedge n_p > n_p') \implies n_s > n_s'$.
\end{definition}
\noindent
For example, in \autoref{fig:bipartite2}, the edges $1, 4, 7$ form a canonical maximum matching while
$2, 3, 7$ do not. All other matchings are not maximal, so this matching yields the only match with
$\sigma = \{ x \mapsto h(a) \}$. For finding the first maximal matching we use the Hopcroft-Karp
algorithm \citep{Hopcroft1973}, but any algorithm for finding a maximum matching can be used. This
initial matching can be found in $\mathcal{O}(m\sqrt{n})$ where $n$ is the number of nodes and
$m$ the number of edges in the bipartite graph. Once an initial maximum matching has been found, we
can apply the algorithm described by Uno, Fukuda and Matsui in \cite{Fukuda1994,Uno1997} to
enumerate all maximum matchings. Each successive matching can be found in $\mathcal{O}(n)$ so in
total we have a time complexity of $\mathcal{O}(m\sqrt{n} + N_m n)$ where $N_m$ is the number of
matchings. For every matching the compatibility and every combination of substitutions in the
matching the compatibility needs to be checked and, if successful, a union substitution is created.

\subsection{Sequence Variables}

So far, we have only considered commutative patterns without sequence variables. In this case,
every maximum matching is also a perfect matching, because the number of
subterms in both pattern and subject must be exactly the same in order for a match to exist. When
we introduce sequence variables into a commutative pattern, this constraint is loosened and we only
need at least as many subterms in the subject as we have non-sequence-variable terms in the pattern.
The remaining subject terms are distributed among the sequence variables.

Consider the pattern $f(g_c(a, x, x, y^*))$, which is one of the previous examples except for an
additional sequence variable $y^*$. However, now it matches the subject $f(g_c(a, a, a, h(a), h(a)))$,
\eg with $\{ x \mapsto h(a), y^* \mapsto \multiset{ a, a } \}$. 
For that purpose, once a maximum matching in the bipartite graph is found, the
remaining unmatched terms are distributed among the sequence variables. If the resulting
substitution is compatible with the one from the matching, the union of the substitutions
yields a valid match. For finding the sequence variable matches, the algorithm discussed in section
\ref{sec:seq_vars} is used.



\chapter{Experimental Results} \label{chp:experiments}

This chapter contains a brief introduction of our implementation of the previously described
algorithms as well as a description of our experiments. Finally, the results of those experiments
are given and discussed.

\section{Python Implementation}

We have implemented the matching algorithms described in the previous chapters in Python. The code
is open source and licensed under the MIT license. It is hosted on Github\footnote{\url{https://github.com/HPAC/matchpy}}.
It has also been made available as the PyPI package \texttt{MatchPy}\footnote{\url{https://pypi.python.org/pypi/matchpy}} for easy
installation for anyone who wants to use it in their Python code. The documentation for the code
can be found on ReadTheDocs\footnote{\url{https://matchpy.readthedocs.io/}}.

The choice to use Python as a language had two main reasons. The first is simply that the
application in which the pattern matching was going to be used was written in Python. The second
reason is a powerful language construct in Python called generators. Generators make writing
code with backtracking easier.

Python does have a disadvantage in terms of speed compared to other languages though: Typically,
python programs are one or two magnitudes slower than programs written in languages such as C or C++
\citep{LanguageBenchmark}. Still, the simpler implementation and easier integration outweighed the
performance loss for us.

\subsection{Generators}

Generators are an extension of the iterator concept which a lot of languages support.
Usually, an iterator is an object that enables a user to loop over a set of values yielded from it.
This is often used to traverse data structures or to programmatically create sequences of values.
The programmer has to manage the state of the iterator object to be able to return the correct
next value upon request.

In Python, generators are special functions or methods, that use the \texttt{yield} keyword instead of
a regular \texttt{return}. Instead of just returning a value to the caller, the state of the
generator function is saved and the execution can resume from the same state once the caller is
done processing the yielded value. This way, multiple values can be ``returned'' iteratively with
additional processing in between.
The main difference compared to iterators in other languages is the management of state. In Python,
the state of the generator function is managed automatically between yields. In contrast, regular
iterators require manual state management by the programmer.

In addition, there is the \texttt{yield from} keyword in Python that allows passing all values
yielded by another generator on to the caller. This enabled us to implement the algorithms described
previously in a way that yields the matches one by one. Therefore, not every match has to be
enumerated if the consumer of the matches decides that no further matches are needed. Generators
also encapsulate the state of each submatching step and make backtracking to the last choice point
in the exploration of the match search space easier.

Another advantage of generators is their lazy evaluation, \ie the next match is only computed if
it is needed. Lazy AC pattern matching has previously been discussed by Belkhir \citep{Belkhir2012}.

\subsection{Terms}

Terms are represented by classes that reflect the tree structure. For each function
symbol in the problem domain, a subclass needs to be created. The class structure allows to store
additional information specific to the problem domain along with the terms. An example for this
would be matrix dimensions in the linear algebra domain.

In addition, we separate two distinct ideas which so far were united in the concept a variable. First, we allow
to give a name to any subpattern and capture whatever it matches in the match substitution.
Secondly, the ability to match anything is implemented in so called \emph{wildcards}. A regular
wildcard can match any term and a sequence wildcard can match a sequence of terms. A named wildcard
is equivalent to a variable. However, unnamed wildcards function as anonymous variables, \ie
their values are not captured in the substitution. Assigning a variable name to terms other than
a wildcard is mostly a convenience feature that allows easier access to complex matched terms
from the substitution.

Furthermore, in some domains, symbols belong to disjoint classes. As an example, in the domain of
linear algebra, symbols can either be matrices, vectors, and scalars. Since these need to be
distinguished by patterns frequently, we decided to make this kind of type constraint a first-class
feature of the pattern matching: Users can create symbol subclasses and use special
\emph{symbol wildcards} to match only symbols of that type.

We always represent terms in their canonical form as described in sections \ref{sec:assoc_func}
and \ref{sec:comm_func}. Hence, term arguments are flattened and sorted on creation if necessary.

\subsection{Constraints}

Patterns can have any number of additional constraints imposed on them. Users can create their own constraint
subclasses or use anonymous functions as constraints. Constraints can either be global, \ie they
are only applied after the whole matching is completed, or they can be local. Local constraints
must define the set of variables upon which they depend, so that they can be evaluated
once all those variables have been assigned a substitution during the matching. This enables early
backtracking in case the constraint fails while freeing the user from having to decide when to
evaluate a constraint themselves. Local constraints are only evaluated once, unless a substitution
for one of their dependency variables changes.

\section{Example Problems}

We carried out experiments to compare the performance of one-to-one and many-to-one pattern matching.
For those experiments, we used two example problems from two different domains. The first
problem is finding subexpressions of linear algebra expressions that match
a kernel, \ie a library function which can compute the result of the subexpression efficiently. The second problem is based on \glspl{AST} of
Python source code. It adapts a set of patterns used in the Python tool \texttt{2to3} which is a
tool used to translate Python source code from version 2 to 3. This translation is necessary, because there were
significant changes between those versions making them source-incompatible.
In the following sections, those problems and their
respective patterns are discussed.

\subsection{Linear Algebra} \label{sec:linalg_experiment}

The terms of the linear algebra problem consist of the function symbols shown in \autoref{tbl:laop}.
Constant symbols can either be scalars, vectors or matrices. In addition, matrices may have
properties like being square, diagonal, triangular, \etc. While matrices and vectors have associated
dimension information, it is not relevant for the pattern matching.

\begin{table}[h]
    \centering
    \renewcommand{\arraystretch}{1.2}
    \begin{tabular}{l c c l}
        \toprule
        Operation & Symbol & Arity & Properties \\
        \midrule
        Multiplication & $\times$ & variadic & associative \\
        Addition & $+$ & variadic & associative, commutative \\
        Transposition & ${}^T$ & unary & \\
        Inversion & ${}^{-1}$ & unary & \\
        Inversion and Transposition & ${}^{-T}$ & unary & \\
        \bottomrule
    \end{tabular}
    \vspace{\baselineskip}
    \caption{Operations for \texttt{LinAlg}}
    \label{tbl:laop}
\end{table}
\noindent
As an example for a pattern, consider \[\alpha \times A \times B + \beta \times C\] which describes
an expression that can be computed using the \texttt{dgemm} routine in the \gls{BLAS}.
Here, $\alpha$ and $\beta$ are scalar variables and $A$, $B$, and $C$ are matrix variables
without special property requirements. Similarly, the pattern
\[A^T \times B \mbox{ \textbf{if} A is square and upper triangular} \]
can be computed by with the \texttt{dtrmm} \gls{BLAS} routine. To allow these pattern to match any
part of a sum or product, respectively, we add sequence variables capturing the rest of the subject term:
\[\alpha \times A \times B + \beta \times C + c^*\]
\[c_1^* \times A^T \times B \times c_2^* \mbox{ \textbf{if} A is square and upper triangular} \]
We call these added variables context variables.

For the experiments, a total of 199 distinct patterns similar to the ones above were used. Out of
those, 61 have an addition as outermost operation, 135 are patterns for products, and 3 are patterns for single matrices.
A lot of these patterns only differ in terms of constraints, \eg there
are 10 distinct patterns matching $A \times B$ with different constraints on the two matrices.
Without the context variables, the product patterns are syntactic except for the associativity of
the multiplication. Because those patterns only contain symbol variables, the associativity does not
influence the matching and we can consider them to be syntactic. In the following, we refer to
the set of patterns with context variables as \texttt{LinAlg}
and the set of syntactic product patterns as \texttt{Syntactic}.

For subjects, we used randomly generated terms with a specific number of operands. For products, each operand has
a probability to either be a scalar (${\sim}16.67\%$), vector (${\sim}27.78\%$) or matrix (${\sim}55.55\%$).
Vectors have a $40\%$ probability of being transposed, while matrices have a $60\%$ probability to be either transposed,
inverted or both. The matrices also have random properties, \ie they can be square, diagonal,
triangular, and symmetric. Note that some of the properties imply or exclude each other, \eg
every diagonal matrix is also triangular. Hence, only valid property combinations were included in
the experiments.

\subsection{Abstract Syntax Trees}

The \texttt{lib2to3} module in python contains ``fixers'' that can translate some aspect of
Python 2 code to Python 3 code. These fixers utilize patterns to discover their targets in the
\gls{AST}. \autoref{lst:ast:pattern} shows an example of such a pattern as it is defined in
the custom language \texttt{lib2to3} uses to express patterns. In that language, quoted strings
denote atomic symbols, names with angle brackets correspond to function symbols, \texttt{any} is a
wildcard and variable names can be assigned with an equal sign. The example pattern in \autoref{lst:ast:pattern} matches a
call to the builtin function \texttt{isinstance} with a tuple as second argument. The fixer is used
to clean up after other fixers, \eg to simplify \texttt{isinstance(x, (int, int))} to
\texttt{isinstance(x, int)}.
\begin{figure}[H]
\begin{lstlisting}[
    language=Python,
    frame=single,
    linewidth=\textwidth,
    xleftmargin=0.1\textwidth,
    xrightmargin=0.1\textwidth
]
power<
    'isinstance'
    trailer< '(' arglist< any ',' atom< '('
        args=testlist_gexp< any+ >
    ')' > > ')' >
>
\end{lstlisting}
\caption{Pattern for \texttt{isinstance} with a tuple of types}
\label{lst:ast:pattern}
\vspace{-\baselineskip}
\end{figure}
To represent this pattern in our notation, we have to define the function symbols
$power, trailer, arglist, atom, testlist\_gexp \in \funcset$. Also, the constant symbols
$s_l, s_r, s_c, isinstance \in \funcset_0$ are used to represent left parenthesis, right parenthesis,
command and \texttt{isinstance}, respectively. Finally, the variables $x \in \varset_0$ and
$y^+ \in \varset_+$ are used. The converted pattern for the above example then becomes
\[power(isinstance, trailer(s_l, arglist(x, s_c, atom(s_l, testlist\_gexp(y^+), s_r)), s_r)).\]
The variables were renamed, but their names are not important for the experiments.

The original patterns support some features thesis not covered in this thesis, \ie features similar to
Mathematica's \texttt{Repeated} and \texttt{Alternatives} (see section~\ref{sec:mathematica}).
Therefore, some of the patterns could not be implemented in our pattern matching framework.
For patterns with alternative subpatterns, a separate pattern for every alternative was
generated. From the original 46 patterns, 36 could be converted. Because of the heavy usage of
nested alternatives, a total of 1223 patterns without alternatives were generated.

Some features of our pattern matcher like commutative function symbols are not used in these
patterns. Nonetheless, the first-class support for symbol types is useful here. \gls{AST} symbols
can have different types, \eg names, string literals, number literals or interpunctuation.
Hence, symbol wildcards are used to distinguish between them.

For subjects, we used 613 examples extracted from the unit tests for the \texttt{lib2to3} fixers.
We concatenated all these example codes into a single file and parsed the \gls{AST} from it. Then
each subexpression in that tree is used as a subject.

The goal of our experiments on this problem is not to perform better than the existing problem
specific implementation in \texttt{lib2to3}, but to show that pattern matching can be adapted to
different domains and to evaluate whether many-to-one matching is beneficial in those cases as
well. We did not compare the speed of our pattern matching and the pattern matching as
implemented in \texttt{lib2to3}, since the features of both are too different for a
meaningful comparison. We will refer to this problem as \texttt{AST}.

\section{Comparisons}

This section compares the performance of various matching algorithms on the problems described in
the previous section. First, we compare one-to-one match and many-to-one matching. Then,
the performance of discrimination nets vs. many-to-one matching on syntactic patterns is compared.
Finally, we evaluate whether Mathematica contains any optimization for many-to-one matching.

All experiment were executed on an Intel Core i5-2500K 3.3 GHz CPU with 8GB of RAM.
Since the algorithms are not parallelized, only one core was used per experiment.
Nonetheless, multiple experiments were run in parallel on the same machine.
The experiments focus on the relative performance of the algorithms rather than absolute performance.

\subsection{One-to-One vs. Many-to-One}

In the first series of experiments, the one-to-one matching algorithm was compared to the many-to-one matching
with regard to the size of the pattern set. A fixed set of subjects was generated, containing 100
terms. Out of those, 30 were sums and 70 were products, matching the distribution in the pattern set.
The number of operands was generated with a normal distribution with $\mu = 5$ and $\sigma = \frac{5}{3}$.
The size of the pattern set was varied by choosing random subsets of the 199 patterns. Across multiple
subsets and repetitions per subject, the mean match and setup times were measured.

\begin{figure}[h]
    \begin{minipage}[b]{.5\linewidth}
        \centering
        \begin{tikzpicture}
            \begin{axis}[
                height=7cm,
                width=\textwidth,
                xlabel={Number of Patterns},
                ylabel={Time [ms]},
                grid=major,
                legend style={at={(0.02,0.97)},anchor=north west},
                xmin=5,
                xmax=195,
                scatter/classes={
                    'm-1'={mark=*,m1plotcolor}
                },
            ]
                \addplot[scatter,only marks,mark options={scale=1.5},scatter src=explicit symbolic,discard if={method}{'1-1'}]
                    table[
                        x=size,
                        y expr=\thisrow{setup_mean}*1000,
                        col sep=comma,
                        meta=method,
                    ] {figures/all_new_patterns_70_30/processed.csv};
            \end{axis}
        \end{tikzpicture}
        \subcaption{Average Setup Time}
        \label{fig:70-30:setup}
    \end{minipage}%
    \begin{minipage}[b]{.5\linewidth}
        \centering
        \begin{tikzpicture}
            \begin{axis}[
                height=7cm,
                width=\textwidth,
                xlabel={Number of Patterns},
                ylabel={Time [ms]},
                grid=major,
                legend style={at={(0.02,0.97)},anchor=north west},
                xmin=5,
                xmax=195,
                scatter/classes={
                    '1-1'={mark=square*,o1plotcolor},
                    'm-1'={mark=*,m1plotcolor}
                },
            ]
                \addplot[scatter,only marks,mark options={scale=1.5},scatter src=explicit symbolic]
                    table[
                        x=size,
                        y expr=\thisrow{match_mean}*1000,
                        col sep=comma,
                        meta=method,
                    ] {figures/all_new_patterns_70_30/processed.csv};
                \legend{{one-to-one}, {many-to-one}}
            \end{axis}
        \end{tikzpicture}
        \subcaption{Average Match Time}
        \label{fig:70-30:match}
    \end{minipage}
    \caption{Times for \texttt{LinAlg}}\label{fig:70-30:times}
\end{figure}

The results for setup time (construction of the \gls{DN}) and match time are shown in
\autoref{fig:70-30:setup} and \autoref{fig:70-30:match}, respectively. The match time is the total
time it takes to match all patterns from the respective pattern set against a single subject. As expected, the construction
time for the many-to-one matcher and its \gls{MLDN} increases with the number of patterns. Likewise,
the match time for one-to-one matching increases linearly with the number of patterns, while
the match time for the many-to-one matcher increases significantly slower. This is also expected,
because the many-to-one matching can exploit similarities between the patterns to match more
efficiently. Therefore, the setup time of the many-to-one matcher and the
match time of the one-to-one matching is interesting, because those are influenced the most by the
increase in pattern set size. For repeated matching against the same pattern set, the setup has to
be done only once and the question is whether it outweighs the matching in terms of time costs.

In \autoref{fig:70-30:speedup}, the speedup of many-to-one matching over one-to-one matching is
shown depending on the pattern set size. In every case, many-to-one matching is faster with a
factor between five and 20. Therefore, when exceeding a certain threshold for the number of subjects, many-to-one
matching is overall faster than one-to-one matching. However, the setup time needs to be amortized
for many-to-one matching to pay off.

\begin{figure}[h]
    \begin{minipage}[b]{.5\linewidth}
        \centering
        \begin{tikzpicture}
            \begin{axis}[
                height=7cm,
                width=\textwidth,
                xlabel={Number of Patterns},
                ylabel=Speedup,
                grid=major,
                legend style={at={(0.02,0.97)},anchor=north west},
                xmin=5,
                xmax=195,
                ymin=0,
            ]
                \addplot[m1plotcolor,only marks,mark options={scale=1.5},mark=*]
                    table[x index=0,y index=1] {figures/all_new_patterns_70_30/var_pattern_count_speedup.dat};
            \end{axis}
        \end{tikzpicture}
        \subcaption{Average Speedup}
        \label{fig:70-30:speedup}
    \end{minipage}%
    \begin{minipage}[b]{.5\linewidth}
        \centering
        \begin{tikzpicture}
            \begin{axis}[
                height=7cm,
                width=\textwidth,
                xlabel={Number of Patterns},
                ylabel={Number of Subjects},
                grid=major,
                legend style={at={(0.02,0.97)},anchor=north west},
                xmin=5,
                xmax=195,
                ymin=0,
            ]
                \addplot[m1plotcolor,only marks,mark options={scale=1.5},mark=*]
                    table[x index=0,y index=1] {figures/all_new_patterns_70_30/var_pattern_count_break_even.dat};
            \end{axis}
        \end{tikzpicture}
        \subcaption{Average Break Even Point}
        \label{fig:70-30:break}
    \end{minipage}
    \caption{Results for \texttt{LinAlg}}\label{fig:70-30}
\end{figure}

The plot in \autoref{fig:70-30:break} shows the break even point for many-to-one matching depending
on the number of patterns. The break even point is the solution $n$ of
\[ n\cdot t_{match\ one-to-one} = n\cdot t_{match\ many-to-one} + t_{setup\ many-to-one}. \]
If the matching is used at least $n$ times, many-to-one matching is faster than one-to-one matching.
Otherwise, the setup time dominates the total time and one-to-one matching is faster.
For example, when matching at least 9 times against the whole pattern set, many-to-one
matching is already faster than one-to-one matching. Every additional match call is then about 18 times
faster than using one-to-one matching.

With the \gls{AST} patterns, similar results were observed. The speedup is even greater
as can be seen in \autoref{fig:ast:speedup}. The significant speedup is most likely due
to most patterns not matching a given subject. While the many-to-one matcher can quickly exclude
most patterns at once, one-to-one matching needs to exclude every pattern individually.
However, as \autoref{fig:ast:break} indicates, the setup time for the underlying \gls{MLDN} is also
much higher, because it has substantially more states. Therefore, the setup time outweighs the
faster matching time until up to about 200 subjects. For the \texttt{2to3} application, the python
files will likely have hundreds of lines resulting in thousands of nodes in the \gls{AST}. Hence
it is faster to use many-to-one matching here in practice.

\begin{figure}[h]
    \begin{minipage}[b]{.5\linewidth}
        \centering
        \begin{tikzpicture}
            \begin{axis}[
                height=7cm,
                width=\textwidth,
                xlabel={Number of Patterns},
                ylabel=Speedup,
                grid=major,
                legend style={at={(0.02,0.97)},anchor=north west},
                xmin=0,
                xmax=1250,
            ]
                \addplot[m1plotcolor,only marks,mark options={scale=1.5},mark=*]
                    table[x index=0,y index=1] {figures/ast/var_pattern_count_speedup.dat};
            \end{axis}
        \end{tikzpicture}
        \subcaption{Average Speedup}
        \label{fig:ast:speedup}
    \end{minipage}%
    \begin{minipage}[b]{.5\linewidth}
        \centering
        \begin{tikzpicture}
            \begin{axis}[
                height=7cm,
                width=\textwidth,
                xlabel={Number of Patterns},
                ylabel={Number of Subjects},
                grid=major,
                legend style={at={(0.02,0.97)},anchor=north west},
                xmin=0,
                xmax=1250,
            ]
                \addplot[m1plotcolor,only marks,mark options={scale=1.5},mark=*]
                    table[x index=0,y index=1] {figures/ast/var_pattern_count_break_even.dat};
            \end{axis}
        \end{tikzpicture}
        \subcaption{Average Break Even Point}
        \label{fig:ast:break}
    \end{minipage}
    \caption{Results for \texttt{AST}}\label{fig:ast}
\end{figure}

\begin{figure}[h]
    \begin{minipage}[b]{.5\linewidth}
        \centering
        \begin{tikzpicture}
            \begin{axis}[
                height=7cm,
                width=\textwidth,
                xlabel={Number of Patterns},
                ylabel={Number of States},
                grid=major,
                legend style={at={(0.02,0.97)},anchor=north west},
                xmin=5,
                xmax=195,
                scatter/classes={
                    'm-1'={mark=*,m1plotcolor}
                },
            ]
                \addplot[scatter,only marks,mark options={scale=1.5},scatter src=explicit symbolic,discard if={method}{'1-1'}]
                    table[
                        x=size,
                        y expr=\thisrow{size_mean},
                        col sep=comma,
                        meta=method,
                    ] {figures/all_new_patterns_70_30/processed.csv};
            \end{axis}
        \end{tikzpicture}
        \subcaption{\texttt{LinAlg}}
        \label{fig:70-30:size}
    \end{minipage}%
    \begin{minipage}[b]{.5\linewidth}
        \centering
        \begin{tikzpicture}
            \begin{axis}[
                height=7cm,
                width=\textwidth,
                xlabel={Number of Patterns},
                ylabel={Number of States},
                grid=major,
                legend style={at={(0.02,0.97)},anchor=north west},
                xmin=0,
                xmax=1250,
                scatter/classes={
                    'm-1'={mark=*,m1plotcolor}
                },
            ]
                \addplot[scatter,only marks,mark options={scale=1.5},scatter src=explicit symbolic,discard if={method}{'1-1'}]
                    table[
                        x=size,
                        y expr=\thisrow{size_mean},
                        col sep=comma,
                        meta=method,
                    ] {figures/ast/processed.csv};
            \end{axis}
        \end{tikzpicture}
        \subcaption{\texttt{AST}}
        \label{fig:ast:size}
    \end{minipage}%
    \caption{Average MLDN Sizes}
        \label{fig:sizes}
\end{figure}
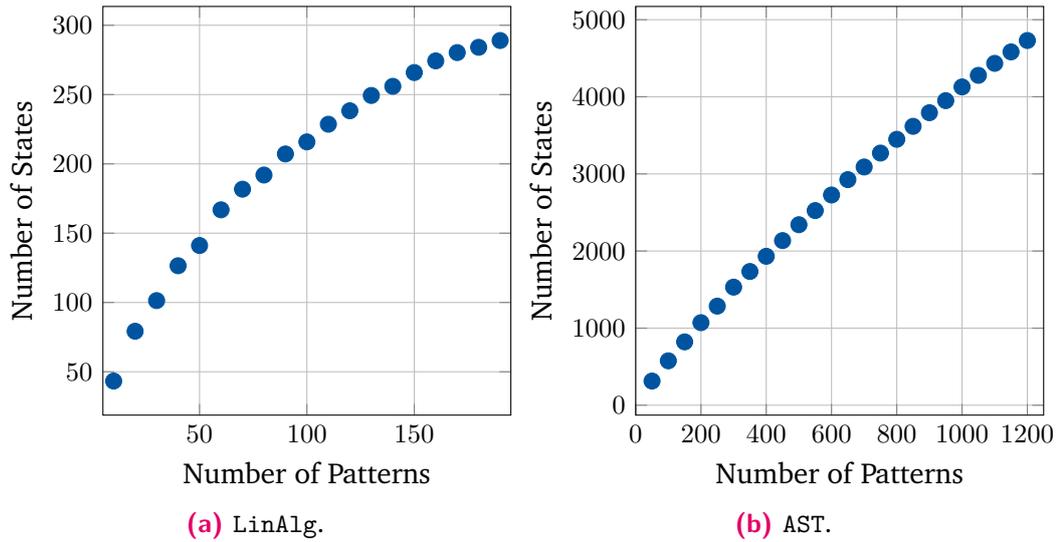

In terms of size, the \gls{MLDN} for the full pattern set of \texttt{LinAlg} has about 300 states,
while the one for all \texttt{AST} patterns has about 5000 states. The growth of the \glspl{MLDN}
is shown in \autoref{fig:sizes}. Because a lot of the patterns for
linear algebra only differ in terms of constraints, the growth slows down with more patterns
as there is more overlap between patterns.

\subsection{Many-to-One vs. Discrimination Net}

In the second set of experiments, our many-to-one matching algorithm was compared with deterministic
discrimination nets for syntactic pattern matching. Again, subsets of the 138 syntactic product
patterns were used and matched against each of a set of 200 subjects. Each subject had between
one and three factors, because that is the range of the pattern operand counts. The factors
were chosen as described in section \ref{sec:linalg_experiment}.

The results are displayed in \autoref{fig:syntactic}. The graph shows that the \gls{DN} is about twice
as fast as the \gls{MLDN}. However, it also has a higher setup time, because for the construction
of the product net is expensive. This is reflected by the break even point for the \gls{VSDN}
which increases even though its speedup also increases
with the number of patterns. Nonetheless, with more than 120 match calls, the deterministic \gls{VSDN}
is the fastest solution, with a speedup of about two over the many-to-one matcher.

\begin{figure}[h]
    \begin{minipage}[b]{.5\linewidth}
        \centering
        \begin{tikzpicture}
            \begin{axis}[
                height=7cm,
                width=\textwidth,
                xlabel={Number of Patterns},
                ylabel=Speedup,
                grid=major,
                legend style={at={(0.02,0.97)},anchor=north west},
                xmin=5,
                xmax=135,
            ]
                \addplot[m1plotcolor,only marks,mark options={scale=1.5},mark=*]
                    table[x index=0,y index=1] {figures/syntactic/var_pattern_count_speedup.dat};
                \addplot[dnplotcolor,only marks,mark options={scale=1.5},mark=triangle*]
                    table[x index=0,y index=2] {figures/syntactic/var_pattern_count_speedup.dat};
                \legend{{MLDN}, {VSDN}}
            \end{axis}
        \end{tikzpicture}
        \subcaption{Speedup}
        \label{fig:syntactic:speedup}
    \end{minipage}%
    \begin{minipage}[b]{.5\linewidth}
        \centering
        \begin{tikzpicture}
            \begin{axis}[
                height=7cm,
                width=\textwidth,
                xlabel={Number of Patterns},
                ylabel={Number of Subjects},
                grid=major,
                legend style={at={(0.02,0.97)},anchor=north west},
                xmin=5,
                xmax=135
            ]
                \addplot[m1plotcolor,only marks,mark options={scale=1.5},mark=*]
                    table[x index=0,y index=1] {figures/syntactic/var_pattern_count_break_even.dat};
                \addplot[dnplotcolor,only marks,mark options={scale=1.5},mark=triangle*]
                    table[x index=0,y index=2] {figures/syntactic/var_pattern_count_break_even.dat};
                \legend{{MLDN}, {VSDN}}
            \end{axis}
        \end{tikzpicture}
        \subcaption{Break Even}
        \label{fig:syntactic:break}
    \end{minipage}
    \caption{Results for \texttt{Syntactic}}
    \label{fig:syntactic}
\end{figure}

In this case, there is no significant
difference in the number of states between \gls{MLDN} and \gls{VSDN}. This is because there is
is not a lot of structural overlap between patterns, \ie if two patterns match the same subject,
the patterns must be equal except for constraints. For example, there is never a situation where
both $x$ and $y^T$ could match the same term (\eg $A^T$), because both variables are symbol
variables that can only match a matrix symbol. For the syntactic subset of patterns from \texttt{AST},
the resulting \gls{VSDN} had 3985 states, while the \gls{MLDN} had 2716 states which is about 32\% less.
Because of the deterministic structure used by the \gls{VSDN}, its size can grow exponentially in
the worst case, while providing a speedup which scales linearly with the number of patterns
in the best case.
Whether that is a useful tradeoff depends on the application and the actual pattern set.

\begin{figure}[h]
    \begin{minipage}[b]{.5\linewidth}
        \centering
        \begin{tikzpicture}
            \begin{axis}[
                height=7cm,
                width=\textwidth,
                xlabel={Number of Patterns},
                ylabel={Number of States},
                grid=major,
                legend style={at={(0.02,0.97)},anchor=north west},
                xmin=5,
                xmax=135,
                scatter/classes={
                    'm-1'={mark=*,m1plotcolor},
                    'DN'={mark=triangle*,dnplotcolor}
                },
            ]
                \addplot[scatter,only marks,mark options={scale=1.5},scatter src=explicit symbolic,discard if not={method}{'DN'}]
                    table[
                        x=size,
                        y expr=\thisrow{size_mean},
                        col sep=comma,
                        meta=method,
                    ] {figures/syntactic/processed.csv};
            \end{axis}
        \end{tikzpicture}
        \subcaption{\texttt{Syntactic}}
        \label{fig:vsdn:size}
    \end{minipage}%
    \begin{minipage}[b]{.5\linewidth}
        \centering
        \begin{tikzpicture}
            \begin{axis}[
                height=7cm,
                width=\textwidth,
                grid=major,
                xlabel={Number of Patterns},
                ylabel={Number of States},
                legend style={at={(0.02,0.97)},anchor=north west},
                ymode=log
            ]
                \addplot[dnplotcolor,only marks,mark options={scale=1.5},mark=triangle*]
                    coordinates {
                        (1, 11)
                        (2, 105)
                        (3, 268)
                        (4, 1974)
                        (5, 3285)
                        (6, 3943)
                        (7, 4393)
                        (8, 34947)
                        (9, 66841)
                        (10, 100429)
                        (11, 124459)
                        (12, 142896)
                        (13, 699956)
                        (14, 729939)
                        (15, 1370213)
                    };
            \end{axis}
        \end{tikzpicture}
        \subcaption{Product Patterns from \texttt{LinAlg}}
        \label{fig:vsdn:size_ctx}
    \end{minipage}
    \caption{Size of VSDN}
\end{figure}
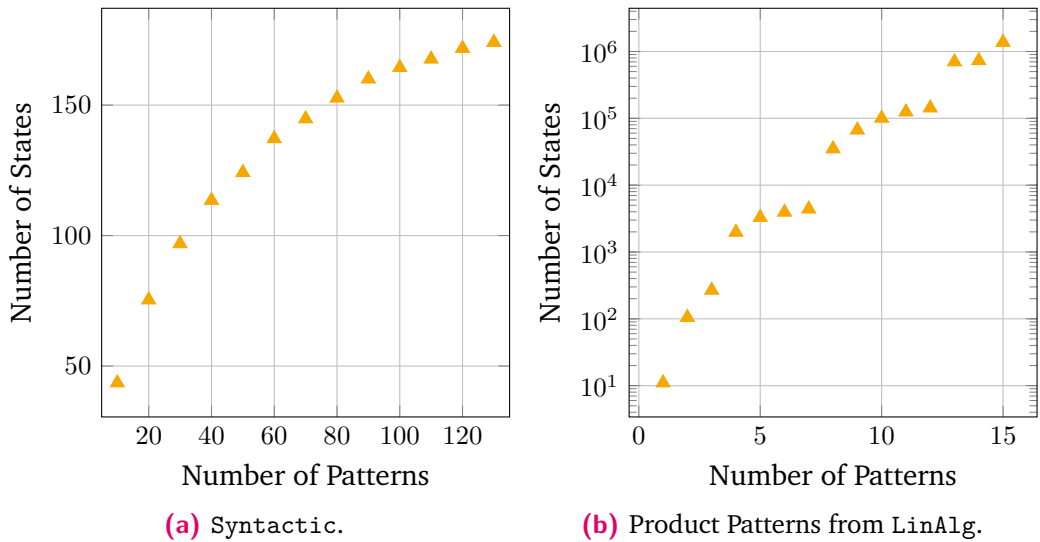

We also tried to use the \gls{VSDN} with patterns that have sequence variables. While it is
possible, the size of the \gls{DN} grows exponentially as illustrated in
\autoref{fig:vsdn:size_ctx}. With just 15 patterns, we have more than a million states already.
Therefore, it is not feasible to use a deterministic \gls{DN} for patterns with sequence variables
that overlap.

\subsection{Mathematica}

To see whether Mathematica implements any kind of optimization for many-to-one matching, we
converted the \texttt{LinAlg} patterns to the Wolfram Language. We prefixed all our custom defined
symbols with ``LinAlg'' to avoid collisions with existing symbols.

For the matrix properties, we use an association that maps a matrix to a list of properties.
An association is similar to a dictionary or hashmap in other languages:
\begin{mmaCell}[index=1]{Code}
LinAlgProperties = <||>;
\end{mmaCell}
We define the properties of the addition and multiplication symbols. Here \texttt{Order\-less}
means commutative, \texttt{Flat} means associative, and \texttt{OneIdentity} means that an operation
with a single operand is equivalent to just that operand.
\begin{mmaCell}{Code}
SetAttributes[LinAlgPlus, {Orderless, OneIdentity, Flat}];
SetAttributes[LinAlgTimes, {OneIdentity, Flat}];
\end{mmaCell}
We define shortcut functions for the symbol types, which we implemented using
\texttt{LinAlgProperties}, \eg every matrix will have the property ``Matrix''.
\begin{mmaCell}[functionlocal={x_,x}]{Code}
LinAlgIsMatrix[x_] := MemberQ[LinAlgProperties[x], "Matrix"]
LinAlgIsVector[x_] := MemberQ[LinAlgProperties[x], "Vector"]
LinAlgIsScalar[x_] := MemberQ[LinAlgProperties[x], "Scalar"]
\end{mmaCell}
Finally, we define the list of patterns as \texttt{LinAlgPatterns}. Every pattern is given as a replacement rule, because
the only way to get a list of all matches in Mathematica is via the \texttt{ReplaceList} function.
As an example, we will use the pattern for the \texttt{dtrmm} \gls{BLAS} routine again:
\[c_1^* \times A^T \times B \times c_2^* \mbox{ \textbf{if} A is square and upper triangular} \]
In Mathematica, this is equivalent to this:
\begin{mmaCell}[functionlocal={ctx1___,A_,B_,ctx2___,ctx1,A,B,ctx2}]{Code}
(LinAlgRoot[LinAlgTimes[ctx1___, LinAlgTranspose[A_?LinAlgIsMatrix], B_?LinAlgIsMatrix, ctx2___]]
   /; SubsetQ[LinAlgProperties[A], {"Square", "UpperTriangular"}])
   -> {54, {{"ctx2", ctx2}, {"A", A}, {"B", B}, {"ctx1", ctx1}}}
\end{mmaCell}
We surround the whole pattern with ``LinAlgRoot'' to get the same behaviour in Mathematica as in our
implementation, \ie that the pattern will only match at the root of the expression, not
anywhere. \texttt{SubsetQ} is used to check if the matrix $A$ has the right properties.
The replacement contains a unique id that identifies the pattern (54 here), and the substitution list.

For our subjects, we first set all the properties of the symbols:
\begin{mmaCell}{Code}
AppendTo[LinAlgProperties, a0 -> {"Scalar"}];
AppendTo[LinAlgProperties, M0 -> {"Matrix", "UpperTriangular", "Square"}];
...
\end{mmaCell}
Then, we define a list of subjects in \texttt{LinAlgSubjects}. As an example, a subject could
look like this:
\begin{mmaCell}{Code}
LinAlgRoot[LinAlgTimes[M8, LinAlgTranspose[v3], M7, LinAlgTranspose[M4], M5, v3, LinAlgInverse[M9]]]
\end{mmaCell}
With the list of 199 patterns and 100 subjects, we get a total of 178 matches:
\begin{mmaCell}[functionlocal={s}]{Code}
AllMatches = Map[Function[s, ReplaceList[s, LinAlgPatterns]], LinAlgSubjects];
NumTotalMatches = Length[Flatten[AllMatches, 1]]
\end{mmaCell}
\begin{mmaCell}{Output}
178
\end{mmaCell}
Finally, this matching process can be repeated for different pattern set sizes:
\begin{mmaCell}[functionlocal={n_,n,s}]{Code}
ManyToOneTimingStep[n_][_] := (ps = RandomSample[LinAlgPatterns, n]; Table[First[AbsoluteTiming[ReplaceList[s, ps]]], {s, LinAlgSubjects}]);
ManyToOneResults = Table[{n, Mean[Flatten[Map[ManyToOneTimingStep[n], Range[100]]]]*1000}, {n, 10, Length[LinAlgPatterns], 10}];
\end{mmaCell}
\texttt{ManyToOneTimingStep} creates a table of the matching times for each subject for a
random pattern subset of size $n$. Note that all patterns are passed to \texttt{ReplaceList}
at once, making this the many-to-one case. The timing set is then repeated 100 times for each
subset size and the mean time is calculated. The times are converted to milliseconds to be better
comparable to the other results.

The same process is repeated for one-to-one matching:
\begin{mmaCell}[functionlocal={n_,n,s,p}]{Code}
OneToOneTimingStep[n_][_] := (ps = RandomSample[LinAlgPatterns, n]; Table[Total[Table[First[AbsoluteTiming[ReplaceList[s, {p}]]], {p, ps}]], {s, LinAlgSubjects}]);
OneToOneResults = Table[{n, Mean[Flatten[Map[OneToOneTimingStep[n], Range[100]]]]*1000}, {n, 10, Length[LinAlgPatterns], 10}];
\end{mmaCell}
The main difference lies in \texttt{OneToOneTimingStep} where every pattern is passed to
\texttt{ReplaceList} separately and the total sum of those times is averaged.

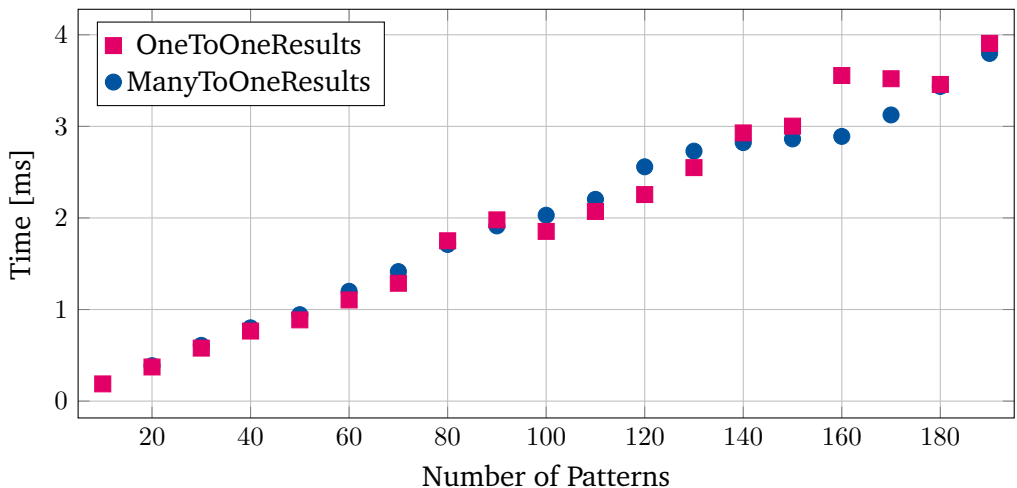
\begin{figure}[h]
    \begin{tikzpicture}
        \begin{axis}[
            height=7cm,
            width=\textwidth,
            xlabel={Number of Patterns},
            ylabel={Time [ms]},
            grid=major,
            legend style={at={(0.02,0.97)},anchor=north west},
            xmin=5,
            xmax=195,
            scatter/classes={
                1-1={mark=square*,o1plotcolor},
                m-1={mark=*,m1plotcolor}
            },
        ]
            \addplot[scatter,only marks,mark options={scale=1.5},scatter src=explicit symbolic]
                table[
                    x index=1,
                    y index=2,
                    col sep=tab,
                    meta index=0
                ] {figures/Mathematica.csv};
			\legend{{OneToOneResults}, {ManyToOneResults}}
        \end{axis}
    \end{tikzpicture}
    \caption{Timing Results for Mathematica}
    \label{fig:mathematica:timing}
\end{figure}

The resulting times are displayed in \autoref{fig:mathematica:timing}. The matching time seems to be
growing linearly with the number of patterns. There is no notable difference between the times of
one-to-one and many-to-one matching. This indicates that there is no many-to-one optimization
implemented in Mathematica.

\chapter{Conclusions and Future Work} \label{chp:conclusions}

\section{Conclusions}

In this thesis we presented algorithms for pattern matching with sequence variables and
associative\Slash{}commutative functions for both one-to-one and many-to-one matching.
Because non-syntactic pattern matching is NP-hard, in the worst case the pattern matching will take
exponential time. Nonetheless, our experiments on real world examples indicate that many-to-one
matching can give a significant speedup over one-to-one matching. However,
the \gls{MLDN} comes with a one-time construction cost which needs to be amortized before it is
faster than one-to-one matching. In our experiments, the break even point for many-to-one
matching was always reached well within the magnitude of the pattern set size. Therefore, for applications that
match more than a few hundred subjects, many-to-one matching can result in a compelling
speedup.

For syntactic patterns, we also compared \glspl{MLDN} with \glspl{VSDN}. As expected,
\glspl{VSDN} are faster at matching, but also have a significantly higher setup time. Furthermore,
their number of states can grow exponentially with the number of patterns, making them unsuitable
for some pattern sets.

Which pattern matching algorithm is the fastest for a given application depends on many
factors. Hence, it is not possible to give a general recommendation. Yet, the more subjects are
matched against the same pattern set, the more likely it is that many-to-one matching pays off.
A higher number of patterns increases the speedup of the many-to-one matching.
In terms of the size of the \gls{MLDN}, the growth of the net seems to be sublinear in practice.
The efficiency of using \glspl{MLDN} also heavily depends on the actual pattern set,
\ie the degree of similarity and overlap between the patterns.

The presented algorithms were implemented in the open-source Python library
\texttt{MatchPy} which is available as a PyPI package.\footnote{\url{https://pypi.python.org/pypi/matchpy}}
We plan on extending the library with more powerful pattern matching features to make it useful for
an even wider range of applications.

\section{Future Work}

There are many possibilities for extending the presented work.  The greatest challenge with
additional features is likely to implement them for many-to-one matching.

\paragraph{Additional pattern features}
In the future, we plan to
implement similar functionality to the \texttt{Repeated}, \texttt{Sequence}, and
\texttt{Alternatives} functions from Mathematica. These provide another level of expressive power
which cannot be replicated with the current feature set of our pattern matching.
Another useful feature are context variables as described by Kutsia \citep{Kutsia2006}.
They allow matching subterms at arbitrary depths which is especially useful for
structures like XML. With context variables, our pattern matching would be as powerful as
XPath \citep{Robie2017} or CSS selectors \citep{Rivoal2017} for such structures.
Similarly, function variables that can match any function symbol would also
be useful for those applications.

\paragraph{SymPy integration}
Integrating \texttt{MatchPy} with the data structures of SymPy (a popular Python library
for symbolic mathematics) would provide the users of SymPy with more powerful pattern matching tools.
However, in SymPy, associativity and commutativity are implemented differently:
Within a function some types of symbols can commute and other do not.
For example, in multiplications, scalars would be commutative while matrices do not commute.
Hence, equivalent functionality would need to be added to \texttt{MatchPy}.

\paragraph{Performance}
Reimplementing the pattern matching in a more performance\hyphen{}oriented language like C or C++ would
likely result in a substantial speedup. While Python is a useful language for prototyping, it
tends to be significantly slower than compiled languages. Alternatively, adapting the code to Cython
\citep{Wilbers2009} could be another option to increase the speed.
Furthermore, generating source code for a pattern set similar to parser generators for formal
grammars could improve matching performance.
While code generation for syntactic pattern matching has been the subject of various works
\citep{Augustsson1985,Fessant2001,Maranget2008,Moreau2003}, its application with the extended
feature set presented in this thesis is another potential area of future research.

\paragraph{Functional pattern matching}
Since Python does not have pattern matching as a language feature, \texttt{MatchPy} could be
extended to provide a syntax similar to other functional programming languages.
Furthermore, in addition to custom symbols, \texttt{MatchPy} could be extended to match on native structures such as tuples
and lists inherently. This would permit writing code in a manner closer to functional languages.

{%
\setstretch{1.1}
\renewcommand{\bibfont}{\normalfont\small}
\setlength{\biblabelsep}{0.3em}
\setlength{\bibitemsep}{0.5\baselineskip plus 0.5\baselineskip}
\Urlmuskip=0mu plus 1mu\relax
\def\UrlBreaks{\do\/\do-}
\printbibliography
}
\cleardoublepage

\listoffigures
\cleardoublepage


\renewcommand\listtheoremname{List of Definitions}
\listoftheorems[ignoreall,show={definition}]

\glsaddall
\cleardoublepage
\phantomsection
\addcontentsline{toc}{chapter}{List of Abbreviations}
\printglossary[type=acronym,title={List of Abbreviations}]


\end{document}